\documentclass[]{elsarticle}
\usepackage[margin=3cm]{geometry}
\usepackage{graphicx}
\usepackage{tikz}
\usepackage{pgf}
\usepackage{commath}
\usepackage{amssymb}
\usepackage{amsmath}
\usepackage{hyperref}
\usepackage{lineno}
\usepackage[english]{babel}
\usepackage[T1]{fontenc}
\usepackage[utf8]{inputenc}

\journal{Computers \& Fluids}

\graphicspath{{./graphicsmake/}}

\definecolor{col1}{HTML}{6988B5}
\definecolor{col2}{HTML}{516168}
\definecolor{col3}{HTML}{945C5B}
\definecolor{col4}{HTML}{BE6C56}
\definecolor{col5}{HTML}{FA9C56}
\definecolor{col6}{HTML}{FEC967}
\definecolor{col7}{HTML}{FFE77D}
\definecolor{col8}{HTML}{CFCE3C}

\renewcommand{\r}[1]{\mathrm{#1}}
\renewcommand{\b}[1]{\mathbf{#1}}
\newcommand{\bs}[1]{\boldsymbol{#1}}
\newcommand{\bt}[1]{\mathbf{#1}}
\renewcommand{\o}[1]{\overline{#1}}
\newcommand{\loc}[1]{#1'}

\begin{document}

\begin{frontmatter}

\title{A finite area scheme for shallow granular flows on three-dimensional surfaces}

\author[igt,bfw]{M. Rauter\corref{ca}}
\ead{matthias.rauter@uibk.ac.at}

\cortext[ca]{Corresponding author address:\\%
Matthias Rauter\\%
University of Innsbruck, Institute of Infrastructure, Division of Geotechnical and Tunnel Engineering,\\%
A-6020 Innsbruck, Austria; Tel.:  +43 (0)512 / 507-62381}

\author[fsb]{\v{Z}. Tukovi{\'c}}
\ead{Zeljko.Tukovic@fsb.hr}

\address[igt]{University of Innsbruck, Institute of Infrastructure, Division of Geotechnical and Tunnel Engineering}
\address[bfw]{Department of Natural Hazards, Austrian Research Centre for Forests (BFW), Innsbruck, Austria}
\address[fsb]{University of Zagreb, Faculty of Mechanical Engineering and Naval Architecture}

\begin{abstract}
Shallow flow or thin liquid film models are used for a wide range of physical and engineering problems. They are applicable to free surface flows, when the flow thickness is distinctly smaller than the lateral dimension. Respective situations exist in automotive and aircraft engineering (soiling, icing, exterior water management), industrial processes (coating, fire suppression), environmental flows (river and lake hydrodynamics, atmospheric flows), and natural hazards (floods, avalanches, debris flow). Shallow flow models allow capturing the free surface of the fluid with little effort and reducing the three-dimensional problem to a quasi two-dimensional problem through depth-integrating the flow fields. Despite remarkable progress of such models in the last decade, accurate description of complex topography remains a challenge. Interaction with topography is particularly critical for granular flows, because their rheology requires modeling of the pressure field, which is strongly linked to surface curvature and associated centrifugal forces. Shallow granular flow models are usually set up in surface-aligned curvilinear coordinates, and velocity is represented as a two-dimensional surface-aligned vector field. The transformation from Cartesian to curvilinear coordinates introduces fictitious forces, however, which result in complex governing equations. In this paper, we set up the shallow flow model in three-dimensional Cartesian coordinates and preserve three-dimensional velocity in the depth-integrated model. Topography is taken into account with a constraint on velocity. This approach is commonly applied by the thin liquid film community. The advantage is a curvature-free mathematical description that is convenient for complex topographies. The constraint on velocity yields a solution for the pressure field, which is required for the pressure-dependent rheology of granular materials. The model is therefore well-suited for granular flows on three-dimensional terrain, e.g., avalanches. The mathematical model is solved with a second-order accurate, implicit finite area scheme, based on the open source software OpenFOAM. We conduct numerical simulations for various cases to verify the numerical routine based on comparisons with analytical results and a grid refinement study. We establish connections to former solutions and discuss advantages and drawbacks of the proposed method from both mechanical and numerical perspectives. 
\end{abstract}


\end{frontmatter}

\section{Introduction}

The first version of the Shallow Water Equations was introduced by Barr\'{e} de Saint-Venant \cite{saint1871theorie}. Since then, they have been extended and adapted to a wide range of problems. The Shallow Water Equations describe free surface flow on an approximately flat surface. This model can be applied when flow thickness is distinctly smaller than the lateral extension of the flow and the fluid velocity can be considered horizontal. The basic idea of the Shallow Water Equations is depth-integration. Vertical variations of field quantities, i.e., pressure and velocity, are assumed to follow a predefined shape function, and average values are used for the mathematical description. In addition, three-dimensional conservation equations are reduced to two-dimensional ones. Moreover, by default, the Shallow Water Equations contain a tracking mechanism for the free surface. In the field of gravitational mass flows, e.g., avalanches, landslides and debris flows, a special case of the Shallow Water Equations, the so-called Savage-Hutter model \citep{gregorian1967new, savage1989motion, savage1991dynamics} is widely utilized \cite{pudasaini2007avalanche}. In automotive and aircraft engineering, this flow configuration, again slightly modified, is known as the thin liquid film model \citep{craster2009dynamics, hagemeier2011practice}. It is applied to simulate the flow of rain water along the vehicle surface. The same holds for fire suppression simulations \citep{xin2011fire} and coating processes \citep{kim2009flow, vita2012thin}.

The granular flow and thin film communities differ in what they deem curvature effects. Thin films of fluid experience a pressure jump at the free surface due to surface tension, depending on the curvature of the free surface (e.g., \cite[][]{moriarty1991unsteady}). Fast granular flows, on the other hand, experience a substantial change in pressure due to centrifugal forces, when they pass over curved surfaces (e.g., \cite[][]{fischer2012topographic}). This work focuses on the latter, and aims to describe the flow of granular matter on a curved three-dimensional surface. 

\begin{figure}
\centering
\includegraphics{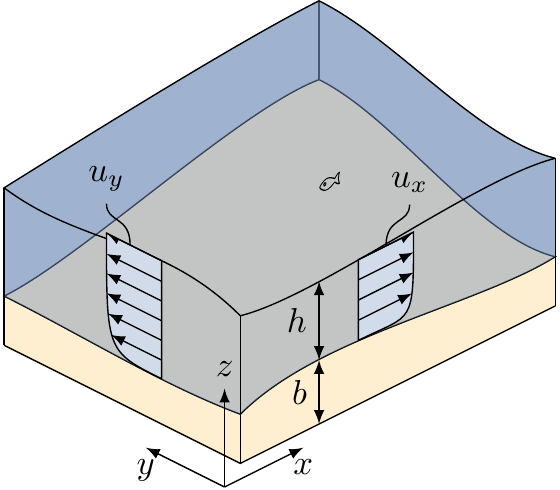}
\caption{Shallow water velocity setting.}
\label{fig:3Dswe}
\end{figure}

The classic form of the Shallow Water Equations,\footnote{Multiplications between vectors represent the outer product $\o{\b{u}}\,\o{\b{u}} = \o{\b{u}}\otimes\o{\b{u}} = \o{u}_{i}\,\o{u}_{j}$} describing the system shown in Fig.~\ref{fig:3Dswe}, is given by:
\begin{eqnarray}
&\dfrac{\partial\,h}{\partial t} + \bs{\nabla} \bs{\cdot} \left(h\,\o{\b{u}}\right) = 0,\label{eq:classic_mass}\\
&\dfrac{\partial\,\left(h\,\o{\b{u}}\right)}{\partial t} + \bs{\nabla} \bs{\cdot} \left(h\,\o{\b{u}}\,\o{\b{u}}\right) =
\dfrac{1}{\rho}\,\bs{\tau}_b-g\,h\,\bs{\nabla}\,b-\dfrac{1}{2}\,g\,\bs{\nabla}\,h^2,\label{eq:classic_mom}
\end{eqnarray}
where $h$ is the vertical flow thickness; $\o{\b{u}} = \left(\o{u}_x, \o{u}_y\right)^\r{T}$ is the depth-averaged horizontal fluid velocity; $g$ is the gravitational acceleration; $\bs{\tau}_{\r{b}}$ is a viscous or basal friction term, usually expressed with the empiric Manning-Strickler equation \citep{manning1891openchannel}; and $b$ is the surface elevation (cf. \cite[][]{vreugdenhil1994numerical}). Eq.~\eqref{eq:classic_mass} represents the depth-integrated continuity equation, and Eq.~\eqref{eq:classic_mom} is the depth-integrated momentum conservation. Topography is taken into account with the second term on the right-hand-side of Eq.~\eqref{eq:classic_mom}. The last term of Eq.~\eqref{eq:classic_mom} expresses the depth-integrated hydrostatic pressure gradient. The direction of the velocity is not linked to the terrain inclination $\bs{\nabla}\,b$ and is considered horizontal, even if the terrain is inclined. This is reasonable on approximately flat terrains, where the vertical component of the velocity is neglectable in comparison to horizontal components.

\begin{figure}
\centering
\includegraphics{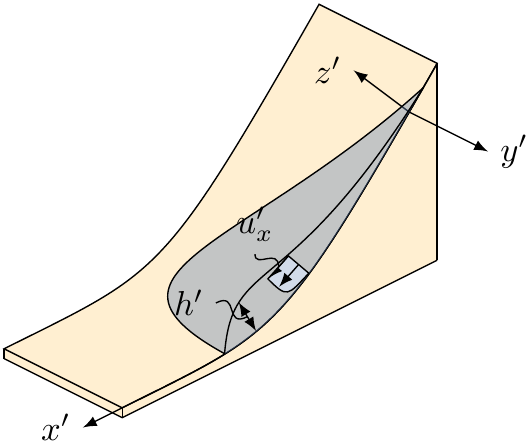}
\caption{Setting for the Savage Hutter model.}
\label{fig:3Dsh}
\end{figure}

Granular flows are always associated with considerable inclination, required to overcome the initial solid-like behavior and the high basal friction. Therefore, the Shallow Water Equations are not applicable without modification. A formal derivation of the Shallow Water Equations for granular flows was first presented by Savage and Hutter \cite{savage1989motion, savage1991dynamics}, while the first application is attributed to Grigorian et al.\ \cite{gregorian1967new}. Savage and Hutter extended the rheological model with concepts from soil mechanics and introduced surface-aligned, curvilinear coordinates $\loc{x}$-$\loc{y}$-$\loc{z}$ (see Fig.~\ref{fig:3Dsh}) to account for the inclined topography associated with granular flows. As a result, the velocity is surface-tangential, rather than horizontal. The Savage-Hutter model, extended by Greve et al.\ \cite{greve1994unconfined} for a simply curved surface, as shown in Fig.~\ref{fig:3Dsh}, reads:
\begin{eqnarray}
&\dfrac{\partial\,\loc{h}}{\partial t} + \loc{\bs{\nabla}}\bs{\cdot} \left(\loc{h}\,\loc{\o{\b{u}}}\right) = 0,\label{eq:sh_mass}\\
&\dfrac{\partial\,\left(\loc{h}\,\loc{\o{\b{u}}}\right)}{\partial t} + \loc{\bs{\nabla}}\bs{\cdot} \left(\loc{h}\,\loc{\o{\b{u}}}\,\loc{\o{\b{u}}}\right) =
\dfrac{1}{\rho}\,\loc{\bs{\tau}_b}+\loc{\b{g}}_{xy}\,\loc{h}-\dfrac{1}{2\,\rho}\,k\,\loc{\bs{\nabla}}\,\left(\loc{h}\,p_{\r{b}}\right).\label{eq:sh_mom}
\end{eqnarray}
Variables denoted with prime refer to curvilinear coordinates, e.g., $\loc{h}$ represents the flow thickness normal to the surface (e.g., in the $\loc{z}$-direction), $\loc{\o{\b{u}}} = (\loc{\o{u}}_x, \loc{\o{u}}_y)^\r{T}$ the fluid velocity in curvilinear coordinates, and $\loc{\bs{\nabla}} = \left(\frac{\partial}{\partial \loc{x}}, \frac{\partial}{\partial \loc{y}}\right)^\r{T}$ the spatial derivatives in curvilinear coordinates. 
\begin{equation}
p_{\r{b}} = \underbrace{\rho\,\loc{h}\,\loc{g}_{z}}_{\mathrm{gravity}}+\underbrace{\rho\,\loc{h}\,\kappa\,{\loc{\o{u}}_x}^2}_{\mathrm{centrifugal\, force}}\label{eq:classic_kappa}
\end{equation}
is the basal pressure, following from gravitational acceleration and centrifugal forces due to the curvature of the ground. $\kappa$ is the curvature of the $\loc{x}$-axis. The basal friction term $\loc{\bs{\tau}}_{\r{b}}$ was initially based on Coulomb friction, but various extensions have been proposed, incorporating influences of velocity (e.g.,~the Voellmy model \cite{voellmy1955uber15}), flow thickness, and granular temperature (e.g., \citep{norem1987continuum, pouliquen2002friction, issler2008exploring, buser2009production, rauter2016snow}). Moreover, the coefficient $k$ takes into account the anisotropic stress state known from soil mechanics. Studies on rough surfaces suggest that the stress state is nearly isotropic in granular flows, and thus $k = 1$ (e.g., \cite[][]{silbert2001granular, pouliquen2002friction, gray2003shock}). This behavior can be linked to the formation of shearing over the whole flow thickness and no slip at the basal surface, whereas Savage and Hutter assumed a plug flow, i.e., slip at the basal surface. The gravitational acceleration is split into a surface-tangential component $\loc{\b{g}}_{xy} = (\loc{g}_x, \loc{g}_y)^\r{T}$ and surface-normal component $\loc{g}_z$. 

Fig.~\ref{fig:3Dsh} shows a special case, in which the surface is only curved along the central flow path, matching the curvilinear $\loc{x}$-axis. Generally, curved topographies are not suitable to set up a curvilinear coordinate system as required by the Savage-Hutter model. Therefore, further extensions have been proposed by introducing a superimposed shallow terrain \citep{gray1999gravity}, a twisted central flow path \citep{pudasaini2005rapid}, and curvature effects on friction \citep{fischer2012topographic}. In addition, a description in Cartesian coordinates, similar to the Shallow Water Equations, has been investigated \citep{bouchut2004gravity, denlinger2004granular, hergarten2015modelling}, taking into account the inclination with various correction terms.

Despite the problems linked to complex topography, the Savage-Hutter model and its extensions have been successfully employed to predict the runout of avalanches, landslides, debris flow, and similar catastrophic events (e.g. \citep[][]{sampl2004avalanche, pitman2003computing, christen2010ramms}). In practice, complex terrain is considered in a simplified manner. SamosAT \cite{sampl2004avalanche} takes advantage of the Lagrangian approach to incorporate complex terrain. Titan2D \cite{pitman2003computing} and Avaflow \cite{mergili2012physically} approximate terrain with an inclined plane and take the difference into account with a correction term, similar to the second term on the right hand side of Eq.~\eqref{eq:classic_mom} \citep{gray1999gravity}. RAMMS applies non-orthogonal local coordinate systems \cite{fischer2012topographic}, however, without incorporation of respective metric tensors \cite{pudasaini2003gravity,hergarten2015modelling}. All simplifications yield model errors, which might become substantial if assumptions are not fulfilled or if approximations are unjustified.

To solve the governing two-dimensional equations, finite difference schemes (e.g., \cite{wang2004savage}), smoothed particle hydrodynamics (e.g., \citep[][]{sampl2004avalanche}), or two-dimensional finite volume methods (e.g., \citep[][]{christen2010ramms}) are applied.

In this work, we follow the general idea of Denlinger and Iverson \cite{denlinger2004granular} and Bouchut et al. \citep{bouchut2004gravity}, and set up the model equations in a Cartesian coordinate system. This corresponds to the usual approach of the thin liquid film community (e.g., \cite{hagemeier2011practice}). The inclination of the topography is considered by incorporating the full three-dimensional depth-integrated velocity vector. Previous works \cite{bouchut2004gravity, denlinger2004granular} utilize the topology constraint to algebraically eliminate the vertical component of the velocity, and the resulting equations can be solved on a horizontal two-dimensional Cartesian grid. In contrast, we use this constraint to calculate basal pressure. Basal pressure is reintroduced into the momentum equation to obtain a solution for velocity. This results in a surface partial differential equation (PDE), similar to other advective-diffusive transport equations on surfaces, such as thin liquid film (e.g., \cite[][]{craster2009dynamics}), surfactant transport (e.g., \cite[][]{dieter2015numerical, olshanskii2010finite, xu2003eulerian, tukovic2012moving}), or acoustic wave propagation (e.g., \cite{lubich2015variational}). For a general review of surface PDEs and additional applications, we refer to Deckelnick et al. \cite{deckelnick2005computation}. With this formulation, all information about topography, such as inclination and curvature, is included in the surface mesh. The governing equations take a very simple form, similar to the basic Shallow Water Equations, without the need for further correction terms. Hence, the method allows to simulate avalanches on complex terrain without further correction terms or metric tensors.

In terms of rheology, we follow the assumption of rough surfaces and consequently shear-dominated deformation in the whole fluid domain. This approach has proven to be useful considering real scale avalanches \cite{issler2008exploring, rauter2016snow}. Moreover, it leads to a very simple implementation of rheology since lateral deviatoric stress can be neglected in shear-dominated flows (e.g., \cite[][]{silbert2001granular, pouliquen2002friction}). The flow thickness is assumed to be much smaller than the curvature radius of the topography, and effects of curvature on depth-integration (cf. \cite[][]{bouchut2003new}) are outside of the scope of this work. Therefore, the proposed method is limited to mildly curved surfaces. The governing equations are solved using the finite area method, i.e., a finite volume method for curved surfaces in three-dimensional space \cite{tukovic2012moving}. We apply an implicit time integration scheme (e.g., \cite[][]{moriarty1991unsteady}). However, the time integration scheme is exchangeable and not a central part of this paper.

The paper is organized as follows: In section \ref{sec:math}, the mathematical derivation of the shallow flow model is presented. For some readers, this may seem to constitute a repetition of classic derivations (e.g., \cite[][]{savage1989motion, savage1991dynamics, greve1994unconfined, gray1999gravity}). However, they differ remarkably since velocity is three-dimensional throughout the whole process. In section \ref{sec:fam}, the discretization on finite areas is explained. Section \ref{sec:implementation} deals with the implementation of the governing equations based on the CFD toolbox OpenFOAM. In section \ref{sec:affinity}, a link to former approaches is established, and some test cases are provided in section \ref{sec:results}. Finally, we present some conclusions and discuss further developments in section \ref{sec:outlook}.

\section{Mathematical description}
\label{sec:math}

To derive a mathematical model for shallow flows on a mildly curved surface, we apply a slightly modified version of the boundary layer approximation \cite[p. 15]{ferziger2002computational}. The assumptions are broadly similar to the Savage-Hutter model:
\begin{itemize}
\item Velocity in the surface-normal direction is negligible; velocity is approximately surface-tangential. 
\item Surface-tangential derivatives of velocity are negligible in comparison to surface-normal derivatives (shear-dominated flow).
\item Viscous momentum transport in surface-tangential directions is negligible in comparison to convective transport.
\end{itemize}

These assumptions are strongly linked to each other. For small velocity derivatives, the vertical growth or shrink of a control volume, and therefore the velocity in surface-normal direction, are small. Viscous momentum transport is linked to the velocity gradient by the rheological model. This connects the first and last assumption. All assumptions can be validated theoretically with a scaling analysis (e.g.\, \cite[][]{greve1994unconfined, gray2014depth}). Here we rely on former results and presuppose that these assumptions are valid.

The classic boundary layer approximation (cf. \cite[][p. 15]{ferziger2002computational}) assumes constant pressure along the flow thickness. This does not hold for granular flows, however, where effective pressure is zero at the free surface and increases substantially with depth. Therefore, the boundary layer approximation is modified by accounting for the variation of pressure along the flow thickness.

The surface-normal variations of pressure and velocity are assumed to follow a predefined shape function, and depth-integration is applied to derive a model, similar to the well-known Shallow Water Equations (or to the Savage-Hutter model). The model is set up in the global Cartesian coordinate system $x$-$y$-$z$, but depth-integration is performed in the surface-normal direction $\loc{z}$ (see Fig.~\ref{fig:coords}). Consequently, flow thickness $\loc{h}$ is measured normal to the surface.

\subsection{Mass conservation}

\begin{figure}
\centering
\includegraphics{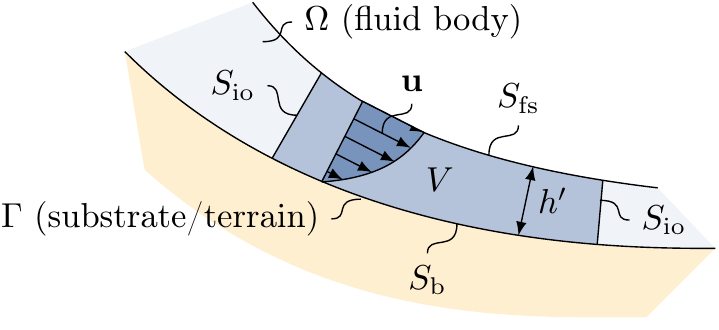}
\caption{Shallow flow configuration: Cut through the control volume.}
\label{fig:cv}
\end{figure}

\begin{figure}
\centering
\includegraphics{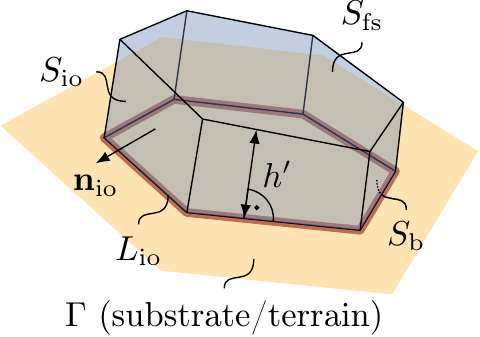}
\caption{Shallow flow configuration: Control volume in three-dimensional space. The boundary line $L_{\r{io}}$, enclosing the surface $S_{\r{b}}$, used to express the area $S_{\r{io}}\approx L_{\r{io}}\,\loc{h}$, is highlighted in red. Edges are drawn as straight lines for simplicity.}
\label{fig:3Dcv}
\end{figure}

\begin{figure}
\centering
\includegraphics{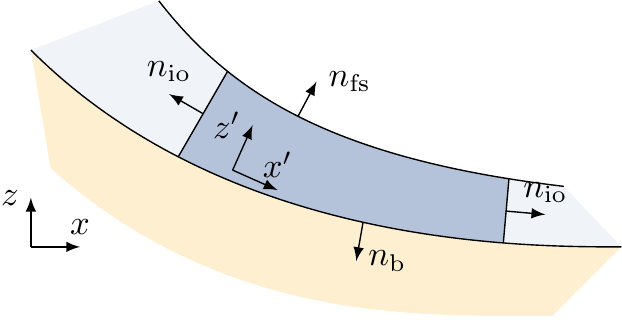}
\caption{Normal vectors.}
\label{fig:coords}
\end{figure}

The conservation of mass for a control volume $V$ within the fluid body $\Omega$ (see Fig.~\ref{fig:cv}) can be written as:
\begin{equation}
\dfrac{\partial}{\partial t}\,\int\limits_{V} \r{d}V+\oint\limits_{S}\b{n} \bs{\cdot}\left(\b{u}-\b{u}_{\r{s}}\right) \, \r{d}S = 0,\label{eq:mass01}
\end{equation}
following Reynolds' transport theorem \citep[pp. 3--6]{ferziger2002computational}. Here, $S=\partial V$ is the boundary of control volume $V$; $\b{u}(\b{x}) \in \mathbb{R}^3, \b{x} \in \Omega \subset \mathbb{R}^3$ (see Fig.~\ref{fig:cv}) is the fluid velocity; $\b{u}_s$ is the velocity of the boundary surface $S$; and $\b{n}$ is the outward-pointing normal vector on $S$. The density is assumed to be constant, and therefore does not appear in this equation. The first term in Eq.~\eqref{eq:mass01} describes the change of the control volume, while the second term describes the net flux through its boundary. To simplify surface fluxes, the boundary surface is split into segments:
\begin{equation}
\dfrac{\partial}{\partial t}\,\int\limits_{V} \r{d}V+\int\limits_{S_{\r{fs}}}\b{n}_{\r{fs}}\bs{\cdot}\left(\b{u}-\b{u}_{\r{fs}}\right) \, \r{d}S
+\int\limits_{S_{\r{b}}}\b{n}_{\r{b}}\bs{\cdot}\left(\b{u}-\b{u}_{\r{b}}\right) \, \r{d}S + \int\limits_{S_{\r{io}}}\b{n}_{\r{io}}\bs{\cdot}(\b{u}-\b{u}_{\r{io}}) \, \r{d}S= 0,\label{eq:mass02}
\end{equation}
where $S_{\r{io}}$ (in and out), $S_{\r{fs}}$ (free surface), and $S_{\r{b}}$ (bottom) are the segments of the boundary surface $S$ (Fig.~\ref{fig:cv}); $\b{n}_{\r{io}}$, $\b{n}_{\r{fs}}$, and $\b{n}_{\r{b}}$ are the respective outward-pointing normal vectors (Fig.~\ref{fig:coords}); and $\b{u}_{\r{io}}$, $\b{u}_{\r{fs}}$, and $\b{u}_{\r{b}}$ are the respective surface velocities. Note that vectors $\b{n}_{\r{io}}$ are usually not normal to $\b{n}_{\r{b}}$, since they are located at different points of the surface. In fact, the angle between $\b{n}_{\r{io}}$ and $\b{n}_{\r{b}}$ carries information about the surface curvature, as shown, e.g., by  Deckelnick et al.\ \cite{deckelnick2005computation}. The upper boundary surface ($S_{\r{fs}}$) moves with the surface-normal velocity of the fluid ($\b{n}_{\r{fs}}\bs{\cdot}\b{u} = \b{n}_{\r{fs}}\bs{\cdot}\b{u}_{\r{fs}}$), so that the control volume captures a whole column of fluid, from the bottom to free surface. Therefore, we can write $\b{n}_{\r{fs}}\bs{\cdot}\left(\b{u} - \b{u}_{\r{fs}}\right) = 0$ and set the flux at the free surface to zero. At the basal surface $S_{\r{b}}$, both liquid velocity $\b{u}$ and boundary velocity $\b{u}_{\r{b}}$ are zero, leading to no flux. The lateral boundaries $S_{\r{io}}$ are stationary, $\b{u}_{\r{io}} = \b{0}$. Eq.~\eqref{eq:mass02} can therefore be written as:
\begin{equation}
\dfrac{\partial}{\partial t}\,\int\limits_{V} \r{d}V+\int\limits_{S_{\r{io}}}\b{n_{\r{io}}}\bs{\cdot} \b{u} \,\r{d}S = 0.\label{eq:mass03}
\end{equation}
To conduct the depth-integration, the volume integral is transformed into a surface-aligned curvilinear coordinate system and split into a double integral:
\begin{equation}
\int\limits_V \r{d}V = \int\limits_{S_{\r{b}}}\int\limits_0^{\loc{h}} \det(\b{J})\,\r{d}\loc{z}\,\r{d}S, \label{eq:vsplit0}
\end{equation}
where $\b{J}$ is the transformation matrix from the global Cartesian coordinate system to the surface-aligned coordinate system. Following the one-dimensional analysis of Bouchut et al. \cite{bouchut2003new}, the determinant can be written as:
\begin{equation}
\det(\b{J}) = 1- \kappa\,\loc{z}. 
\end{equation}
The term $\kappa\,\loc{z}$ can be estimated as the ratio of flow thickness to curvature radius. The definition of shallow flows on mildly curved surfaces, as situations where the flow thickness is considerably smaller than the curvature radius (e.g., \cite{greve1994unconfined}), allows us to simplify $\det(\b{J}) \approx 1$ and
\begin{equation}
\int\limits_V \r{d}V \approx \int\limits_{S_{\r{b}}}\int\limits_0^{\loc{h}} \r{d}\loc{z}\,\r{d}S. \label{eq:vsplit}
\end{equation} 
The same argument can be utilized to split the surface integral over the lateral boundary surface $S_{\r{io}}$:
\begin{equation}
\int\limits_{S_{\r{io}}} \mathrm{d}S = \oint\limits_{L_{\r{io}}}\int\limits_0^{\loc{h}} \det(\b{J})\,\r{d}\loc{z}\,\r{d}L \approx  \oint\limits_{L_{\r{io}}}\int\limits_0^{\loc{h}} \r{d}\loc{z}\,\r{d}L, \label{eq:ssplit}
\end{equation}
where $L_{\r{io}} = \partial S_{\r{b}}$ is the border of surface $S_{\r{b}}$ (see Fig.~\ref{fig:3Dcv}). Denlinger and Iverson \cite{denlinger2004granular} present an alternative approach and employ depth-integration along the global $z$-coordinate, leading to curvature-independent terms for relations~\eqref{eq:vsplit} and \eqref{eq:ssplit}. Introducing all double integrals yields:
\begin{equation}
\dfrac{\partial}{\partial t}\,\int\limits_{S_{\r{b}}} \int\limits_{0}^{\loc{h}} \r{d}\loc{z}\,\r{d}S+\oint\limits_{L_{\r{io}}}\int\limits_{0}^{\loc{h}}\b{n_{\r{io}}}\bs{\cdot} \b{u} \, \r{d}\loc{z} \, \r{d}L = 0.\label{eq:mass04}
\end{equation}
Here we introduced flow thickness $\loc{h}$, as known from other shallow flow models. The evaluation of the first term in Eq.~\eqref{eq:mass04} yields:
\begin{equation}
\loc{h}(\b{x}_{\r{b}}) = \int\limits_{0}^{\loc{h}(\b{x}_{\r{b}})} \r{d}\loc{z},
\end{equation}  
where $\loc{h}(\b{x}_{\r{b}})$ is the surface-normal distance between a point $\b{x}_{\r{b}}$ on the surface $\Gamma$ and the free surface (see Figs.~\ref{fig:cv} and \ref{fig:3Dcv}). The vector $\b{n}_{\r{io}}$ can be taken out from the inner integral of the second term because it is constant along the flow thickness (compare Fig.\ \ref{fig:coords}). The only flow field which has to be integrated is fluid velocity $\b{u}$. To accomplish this, depth-averaged velocity is defined as
\begin{equation}
\o{\b{u}}(\b{x}_{\r{b}}) = \dfrac{1}{\loc{h}(\b{x}_{\r{b}})} \int\limits_{0}^{\loc{h}(\b{x}_{\r{b}})} \b{u}(\b{x}_{\r{b}}-\b{n}_{\r{b}}\,\loc{z})\,\r{d}\loc{z}\label{eq:velocity}
\end{equation}
and introduced in Eq.~\eqref{eq:mass04}. Note that depth-averaged velocity is still a three-dimensional vector field, $\b{\o{u}}(\b{x}_{\r{b}}) \in \mathbb{R}^3, \b{x}_{\r{b}} \in \Gamma \subset \mathbb{R}^3$, defined on the surface $\Gamma \subset \partial \Omega$ (see Fig.~\ref{fig:cv}). Moreover, it has not been transformed to the surface-aligned curvilinear coordinate system as done in the classic approach of Savage and Hutter \cite{savage1989motion, savage1991dynamics}. The final form of the continuity equation in terms of flow thickness $\loc{h}$ and depth-averaged velocity $\o{\b{u}}$ reads
\begin{equation}
\dfrac{\partial}{\partial t} \int\limits_{S_{\r{b}}} \loc{h} \, \r{d}S + \oint\limits_{L_{\r{io}}} \b{n}_{\r{io}}\bs{\cdot} \loc{h}\,\o{\b{u}}\,\r{d}L = 0.\label{eq:mass}
\end{equation}

\subsection{Momentum conservation}

In analogy to Eq.~\eqref{eq:mass01}, the conservation of momentum for control volume $V$ can be written as
\begin{equation}
\dfrac{\partial}{\partial t}\int\limits_{V} \b{u} \,\r{d}V + \oint\limits_{S} \b{n}\bs{\cdot}\b{u}\,\left(\b{u}-\b{u}_s\right)\,\r{d}S
= \dfrac{1}{\rho}\oint\limits_{S} \b{n}\bs{\cdot} \bt{T}^{\r{dev}}\,\r{d}S + \int\limits_{V} \b{g}\,\r{d}V - \dfrac{1}{\rho}\oint\limits_{S}\b{n}\,p\,\r{d}S,\label{eq:momentum01}
\end{equation}
where $\bt{T}^{\r{dev}}$ is the deviatoric or viscous part of the stress tensor; $\b{g}$ is the gravitational acceleration; $\rho$ is the fluid density, assumed to be constant; and $p$ is the pressure. Total stresses can be calculated as $\bt{T} = -p\,\bt{I}+\bt{T}^{\r{dev}}$, where $\bt{I}$ is the identity matrix. Note that compressive stress is negative in $\bt{T}$. Convective fluxes (second term on the left-hand-side) are treated similarly to mass conservation - on surfaces $S_{\r{fs}}$ and $S_{\r{b}}$ the convective flux is zero and on surface $S_{\r{io}}$ the boundary velocity $\b{u}_{\r{io}}$ is zero, leading to:
\begin{equation}
\oint\limits_{S} \b{n}\bs{\cdot}\b{u}\,\left(\b{u}-\b{u}_s\right)\,\r{d}S = \int\limits_{L_{\r{io}}} \int\limits_0^{\loc{h}} \b{n}_{\r{io}}\bs{\cdot}\b{u}\,\b{u}\,\r{d}\loc{z}\,\r{d}L.\label{eq:flux01}
\end{equation}
The  depth-averaged, squared velocity is not equal to the squared, depth-averaged velocity, $ \o{\b{u}\,\b{u}} \neq  \o{\b{u}}\,\o{\b{u}}$. Nevertheless we aim to use depth-averaged velocity to express this term. This is accomplished by introducing shape-factor $\xi$, describing the vertical velocity distribution as:
\begin{equation}
\xi\,\o{\b{u}}\,\o{\b{u}} = \o{\b{u}\,\b{u}} = \dfrac{1}{\loc{h}} \int\limits_{0}^{\loc{h}}\b{u}\,\b{u}\,\r{d}\loc{z}.\label{eq:velocity2}
\end{equation}
For a (vertically) constant velocity profile (plug flow) $\xi=1$, for a linear velocity profile $\xi=4/3$ and for the Bagnold profile (commonly assumed velocity profile in shear-dominated granular flows, see Fig.~\ref{fig:cv}) $\xi = 5/4$, as a few examples \citep{baker2016two}. Recent works (e.g., \cite[][]{richard2016three}) aim to include the varying velocity profile into the shape factor.

\begin{figure}
\centering
\includegraphics{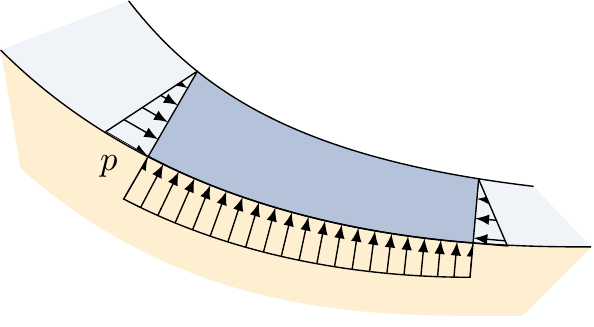}
\caption{Pressure distribution.}
\label{fig:pressure}
\end{figure}

Pressure is assumed to follow a prescribed function $p(\loc{z})$ along the flow thickness. The simplest approach for pressure along the flow thickness is a linear distribution, as shown in Fig.~\ref{fig:pressure}, interpolating between pressure at basal surface $p_{\r{b}}$ and pressure at free surface $p_{\r{fs}} = 0$:
\begin{equation}
p = p_{\r{b}}\,\dfrac{\loc{h}-\loc{z}}{\loc{h}}.\label{eq:pfun}
\end{equation}
This approach is consistent for pressure due to gravity, i.e., hydrostatic or lithostatic pressure. Centrifugal forces, on the other hand, lead to a different pressure profile, depending on surface curvature and velocity profile. However, in a realistic scenario, pressure can be mainly attributed to gravity, and Eq.~\eqref{eq:pfun} seems reasonable. The surface integral of pressure in Eq.~\eqref{eq:momentum01} can be divided into segments $S_{\r{io}}$ and $S_{\r{b}}$. Free surface $S_{\r{fs}}$ is pressure-free and can be neglected. The surface integral over surface $S_{\r{io}}$ can be further split following Eq.~\eqref{eq:ssplit} and depth-integration can be performed analytically:
\begin{equation}
\int\limits_{S_{\r{io}}} \b{n}_{\r{io}}\,p\,\r{d}S = \oint\limits_{L_{\r{io}}}\b{n}_{\r{io}}\int\limits_{0}^{\loc{h}} p\,\r{d}\loc{z}\,\r{d}L = \oint\limits_{L_{\r{io}}}\b{n}_{\r{io}}\int\limits_{0}^{\loc{h}} p_{\r{b}}\,\dfrac{\loc{h}-\loc{z}}{\loc{h}} \,\r{d}\loc{z}\,\r{d}L = \oint\limits_{L_{\r{io}}}\dfrac{1}{2}\,\b{n}_{\r{io}}\,\loc{h}\,p_{\r{b}}\,\r{d}L,\label{eq:pfun2}
\end{equation}
leading to the depth-integrated formulation of the pressure term:
\begin{equation}
\oint\limits_{S}\b{n}\,p\,\r{d}S = \oint\limits_{L_{\r{io}}} \dfrac{1}{2}\,\b{n}_{\r{io}}\,\loc{h}\,p_{\r{b}}\,\r{d}L + \int\limits_{S_{\r{b}}} \b{n}_{\r{b}}\,p_{\r{b}}\,\r{d}S.\label{eq:pressure02}
\end{equation}
Basal pressure $p_{\r{b}}$ is not further described at this point. The boundary condition introduced by topography (section~\ref{ssec:topo}) will yield a result for this field. It is worth noting that the depth-averaged velocity is three-dimensional, and therefore not automatically restricted to follow the terrain. The correct determination of the basal pressure will ensure that velocity remains tangential to topography. The basal pressure will therefore naturally contain all known effects, such as gravity and centrifugal forces. Mathematically, this can be expressed with projections on the surface and the surface-normal vector. This constitutes a key feature of the proposed method and will be discussed further in the following.

The deviatoric stress tensor is expressed by the constitutive model as a function of primary flow fields, i.e., pressure and velocity. Here we apply the $\mu(I)$-rheology \cite{midi2004dense}, which is able to capture the most important aspects of granular material in motion. Following this simple law, the divergence of the deviatoric stress tensor can be replaced with a basal friction term (a detailed scaling analysis can be found in \cite{gray2014depth}):
\begin{equation}
\oint\limits_{S} \b{n}\cdot\b{T}^{\r{dev}}\,\r{d}S = -\int\limits_{S_{\r{b}}} \bs{\tau}_{\r{b}}\,\r{d}{S},\label{eq:basal_friction}
\end{equation}
with the basal friction term
\begin{equation}
\bs{\tau}_{\r{b}} = \mu(I_{\r{b}})\,p_{\r{b}}\dfrac{\b{\o{u}}}{|\b{\o{u}}|+u_0}.\label{eq:muI}
\end{equation}
Friction coefficient $\mu$ is expressed as a function of dimensionless inertial number $I_{\r{b}}$:
\begin{eqnarray}
&&\mu = \mu_{\r{s}} + \dfrac{\mu_2-\mu_{\r{s}}}{I_0/I_{\r{b}}+1},\\
&&I_{\r{b}} = \frac{5}{2}\frac{|\b{\o{u}}|/\loc{h}\,d}{\sqrt{p_{\r{b}}/\rho_{\r{p}}}},
\end{eqnarray}
with the material parameters $d$ (particle diameter), $\rho_{\r{p}}$ (particle density), $\mu_s$, $\mu_2$, and $I_0$. The factor $5/2$ is related to the assumed Bagnold profile.
The additional parameter $u_0$ in the denominator of Eq.~\eqref{eq:muI} regularizes the expression to suppress oscillations near still-stand where the original function is discontinuous \cite{lagree2011granular}. 

Introducing Eqs.~\eqref{eq:flux01}, \eqref{eq:pressure02}, and \eqref{eq:basal_friction} and the depth-averaged entities \eqref{eq:velocity} and \eqref{eq:velocity2} into Eq.~\eqref{eq:momentum01} leads to the final depth-integrated momentum conservation equation:
\begin{eqnarray}
\dfrac{\partial}{\partial t}\int\limits_{S_b} \loc{h}\,\o{\b{u}} \,\r{d}S 
+ \oint\limits_{L_{\r{io}}} \b{n}_{\r{io}}\bs{\cdot}\xi\,\loc{h}\,\o{\b{u}}\,\o{\b{u}}\,\r{d}L
= 
-\dfrac{1}{\rho}\int\limits_{S_{\r{b}}}\bs{\tau}_{\r{b}}\,\r{d}S
+ \int\limits_{S_{\r{b}}} \loc{h}\,\b{g}\,\r{d}S
- \dfrac{1}{2\,\rho}\oint\limits_{L_{\rm{io}}} \,\b{n}_{\r{io}}\,\loc{h}\,p_{\r{b}}\,\r{d}L - \dfrac{1}{\rho}\int\limits_{S_{\r{b}}} \b{n}_{\r{b}}\,p_{\r{b}}\,\r{d}S.\label{eq:momentum0.3}
\end{eqnarray}
Note that the last term in Eq.~\eqref{eq:momentum0.3}, basal pressure, vanishes in the classic derivation, since it is normal to the basal surface (e.g., \cite[][]{savage1989motion, savage1991dynamics}). Here, this term is required to redirect three-dimensional velocity along the surface.

\subsection{Topography boundary condition}
\label{ssec:topo}

The boundary condition
\begin{equation}
\o{\b{u}} \bs{\cdot} \b{n}_{\r{b}} = 0\label{eq:bc}
\end{equation}
ensures that the velocity of the fluid is surface-tangential, as assumed by the boundary layer approximation and the Savage-Hutter model. This constraint can be considered as an equation to determine basal pressure $p_{\r{b}}$ in every point of the surface. In return, basal pressure can be used to calculate the velocity field such that the constraint will be satisfied.

\subsection{Differential form and projection}
\label{ssec:projection}

To solve the flow model, the momentum conservation Eq.~\eqref{eq:momentum0.3} has to be combined with the boundary condition~\eqref{eq:bc}, and split into a surface-tangential and surface-normal part. The surface-tangential part can be solved for velocity, the surface-normal part for basal pressure. In the classic derivation, this occurs naturally due to the curvilinear coordinate system. Here we avoid this coordinate transformation and conduct a projection instead. Since this projection cannot be applied to the integral formulation, we have to rewrite the governing equations in differential form. The mass conservation~\eqref{eq:mass} can be rewritten as:
\begin{equation}
\int\limits_{S_b} \dfrac{\partial\,\loc{h}}{\partial t}\r{d}S 
+ \int\limits_{S_{\r{b}}} \bs{\nabla}\bs{\cdot}\left(\loc{h}\,\o{\b{u}}\right)\,\r{d}S
= 0,
\end{equation}
using the Gauss Theorem and the Leibniz rule. Since this equation holds for any control area $S_{\r{b}}$, the integrand has to be zero \cite{leveque1990conservative},
\begin{equation}
 \dfrac{\partial\,\loc{h}}{\partial t}
+ \bs{\nabla}\bs{\cdot}\left(\loc{h}\,\o{\b{u}}\right)
= 0.\label{eq:differential_mass}
\end{equation}
With the same arguments, the momentum conservation Eq.~\eqref{eq:momentum0.3} can be rewritten as:
\begin{equation}
 \dfrac{\partial\,\loc{h}\,\o{\b{u}}}{\partial t}
+ \xi\,\bs{\nabla}\bs{\cdot}\left(\loc{h}\,\o{\b{u}}\,\o{\b{u}}\right)
= 
-\dfrac{1}{\rho}\bs{\tau}_{\r{b}}
+ \loc{h}\,\b{g}
- \dfrac{1}{2\,\rho}\bs{\nabla}\,\left(\loc{h}\,p_{\r{b}}\right)
- \dfrac{1}{\rho}\b{n}_{\r{b}}\,p_{\r{b}}.\label{eq:differential_total_mom}
\end{equation}
Eq.~\eqref{eq:differential_total_mom} combines the classic momentum conservation equation with the equation for basal pressure. As previously mentioned, this mixed formulation is a consequence of the Cartesian coordinate system. 
To obtain a solution for pressure, we can solve the surface-normal projection of Eq.~\eqref{eq:differential_total_mom}, obtained by multiplying it with the normal vector $\b{n}_{\r{b}}$:
\begin{equation}
 \underbrace{\b{n}_{\r{b}}\bs{\cdot} \dfrac{\partial\,\loc{h}\,\o{\b{u}}}{\partial t}}_{=0}
+ \b{n}_{\r{b}}\bs{\cdot}\xi\,\bs{\nabla}\bs{\cdot}\left(\loc{h}\,\o{\b{u}}\,\o{\b{u}}\right)
= 
- \dfrac{1}{\rho}\,\underbrace{\b{n}_{\r{b}}\bs{\cdot}\bs{\tau}_{\r{b}}}_{=0}
+ \loc{h}\,\b{n}_{\r{b}}\bs{\cdot}\b{g}
- \dfrac{1}{2\,\rho}\b{n}_{\r{b}}\bs{\cdot}\bs{\nabla}\,\left(\loc{h}\,p_{\r{b}}\right)
- \dfrac{1}{\rho}\underbrace{\b{n}_{\r{b}}\bs{\cdot}\b{n}_{\r{b}}}_{=1}\,p_{\r{b}}.\label{eq:differential_normal_mom0}
\end{equation}
The first term yields zero according to the boundary condition, and the basal friction term disappears, as well. An additional multiplication with $\b{n}_{\r{b}}$ yields an equation for pressure, including its direction $\b{n}_{\r{b}}\,p_{\r{b}}$:
\begin{equation}
\xi\,\bs{\nabla}_{\r{n}}\bs{\cdot}\left(\loc{h}\,\o{\b{u}}\,\o{\b{u}}\right)
= \loc{h}\,\b{g}_{\r{n}}
- \dfrac{1}{2\,\rho}\bs{\nabla}_{\r{n}}\,\left(\loc{h}\,p_{\r{b}}\right)
- \dfrac{1}{\rho}\b{n}_{\r{b}}\,p_{\r{b}},\label{eq:differential_normal_mom}
\end{equation}
with the surface-normal gradient operator:
\begin{equation}
\bs{\nabla}_{\r{n}} = \left(\b{n}_{\r{b}}\,\b{n}_{\r{b}}\right)\bs{\cdot}\bs{\nabla},
\end{equation}
and the normal component of gravitational acceleration $\b{g}_{\r{n}} = \left(\b{n}_{\r{b}}\,\b{n}_{\r{b}}\right)\,\b{g}$. The first term in Eq.~\eqref{eq:differential_normal_mom} accounts for centrifugal forces, and the second for gravity. The third term, not considered in the classic approach, takes into account that the gradient of lateral pressure contains a surface-normal component due to surface curvature (compare Fig.~\ref{fig:pressure}). This term is small, but required to obtain a consistently surface-tangential velocity in the next step. However, it plays a minor role in pressure, and consequently basal friction.

Eq.~\eqref{eq:differential_normal_mom} is subtracted from Eq.~\eqref{eq:differential_total_mom} to obtain the surface-tangential momentum equation:
\begin{equation}
 \dfrac{\partial\,\loc{h}\,\o{\b{u}}}{\partial t}
+ \xi\,\bs{\nabla}_{\r{s}}\bs{\cdot}\left(\loc{h}\,\o{\b{u}}\,\o{\b{u}}\right)
= 
- \dfrac{1}{\rho}\bs{\tau}_{\r{b}}
+ \loc{h}\,\b{g}_{\r{s}}
- \dfrac{1}{2\,\rho}\bs{\nabla}_{\r{s}}\,\left(\loc{h}\,p_{\r{b}}\right),\label{eq:differential_tang_mom}
\end{equation}
similar to the momentum conservation of the Shallow Water Equations. The surface-tangential gradient operator:
\begin{equation}
\bs{\nabla}_{\r{s}} = \left(\b{I}-\b{n}_{\r{b}}\,\b{n}_{\r{b}}\right)\bs{\cdot}\bs{\nabla},
\end{equation}
appears in various physical problems on surfaces (e.g., \cite[][]{xu2003eulerian, dziuk2007finite, dziuk2013, tukovic2012moving, lubich2013backward, lubich2015variational, dieter2015numerical}). The surface-normal counter part $\bs{\nabla}_{\r{n}}$ is less popular. A similar projection has previously been utilized to split surface tension into surface-normal and tangential components \cite{tukovic2012moving}. In the context of shallow granular flows, it is a convenient tool to derive a solution for basal pressure.

The advantage of this method is that no artifacts of coordinate transformation appear in the equations. These terms usually contain surface curvature, which is difficult to generalize to complexly curved terrain. Moreover, the equation for basal pressure~\eqref{eq:differential_normal_mom} is similar to the equation for velocity~\eqref{eq:differential_tang_mom}. All terms, except strictly surface-normal or tangential ones, appear in both equations. This highlights the high similarity between velocity and pressure equations, and their common origin. This concept to describe curvature can be extended to any flow model (e.g., two-phase models or anisotropic stress states).

\section{Finite area discretization}
\label{sec:fam}

To numerically solve the flow model, expressed by Eqs.~\eqref{eq:differential_mass}, \eqref{eq:differential_normal_mom} and \eqref{eq:differential_tang_mom}, the computational domain (time and space) is discretized. 
The temporal domain (i.e., the simulation time) is split into a finite number of time steps, and the equations are solved in a time-marching manner. 
The spatial domain, herein a curved surface $\Gamma$ in three-dimensional space, is divided into a finite number of flat and convex polygonal control areas, bounded by an arbitrary number of straight edges. The control areas cover the spatial domain completely without overlapping. 

In this paper, the governing equations are solved implicitly and sequentially but an explicit or coupled approach is also viable. Explicit solvers evaluate all terms using known field values, i.e., values of old time steps, while implicit solvers express all or most terms as functions of unknown field values, i.e., values of the new time step \cite[pp. 142--151]{ferziger2002computational}. The sequential approach solves one governing equation at a time, and each equation yields a result for a single flow field (herein velocity, basal pressure, or flow thickness). The sequential approach was chosen because of the complex nonlinear coupling of flow thickness and velocity. For linear and tightly coupled equations, the simultaneous approach, solving all governing equations together, is usually preferred \cite[p. 117]{ferziger2002computational}.\footnote{Indices in $\bs{\tau}_{\r{b}}$, $p_{\r{b}}$, and $\b{n}_{\r{b}}$ are omitted here for increased readability. The same holds for surface-normal flow thickness $\loc{h}$, which is expressed as $h$ in the following.}

\subsection{Spatial discretization}
\label{sec:spatial}

\begin{figure}
\centering
\includegraphics{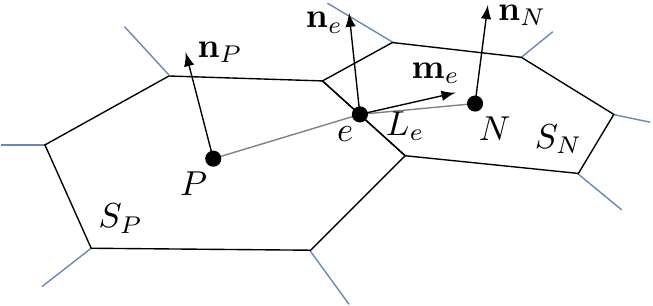}
\caption{Polygonal control areas.}
\label{fig:ca}
\end{figure}

The finite area discretization is based on the integral form of the conservation equations which can be obtained by integrating the governing equations over a control area. Fig.~\ref{fig:ca} presents two polygonal control areas $S_P$ and $S_N$ with computational points $P$ and $N$ at the respective centroids. Edges with length $L_e$ are denoted with $e$. Surface-normal vectors are denoted with $\b{n}_P$ and $\b{n}_N$ for areas and with $\b{n}_e$ for edges. Normal vectors $\b{n}_{\r{io}}$ on lateral surfaces $S_{\r{io}}$ now take the form of bi-normal vectors on edge $e$ and normal vector $\b{n}_e$, and are denoted as $\b{m}_e$.
Using the second-order accurate midpoint rule to express integrals over the surface $S_{\r{b}}$ for cell $P$ yields \cite[pp. 72--74]{ferziger2002computational}:
\begin{equation}
\int\limits_{S_{\r{b}}} \phi \, \r{d}S \approx \phi_{P} \,S_{P},
\end{equation}
where $\phi_{P}$ is the value of field $\phi$ (in the following, $\phi$ is used as a placeholder for arbitrary fields) in computational point $P$.

The gradient and divergence operators are rewritten as boundary integrals over $L_{\r{io}} = \partial S_{\r{b}}$ using the Gauss Theorem, and discretized as:
\begin{eqnarray}
&&\int\limits_{S_{\r{b}}} \bs{\nabla}\,\phi\, \r{d}S = \oint\limits_{L_{\r{io}}} \bs{n}_{\r{io}}\, \phi \, \r{d}L \approx \sum\limits_e \b{m}_{\r{e}}\,\phi_{e}\,L_{e},\label{eq:grad_disc}\\
&&\int\limits_{S_{\r{b}}} \bs{\nabla}\bs{\cdot}\bs\phi\, \r{d}S = \oint\limits_{L_{\r{io}}} \bs{n}_{\r{io}}\bs{\cdot} \bs\phi \, \r{d}L \approx \sum\limits_e \b{m}_{\r{e}} \bs{\cdot} \bs\phi_{e} \,L_{e},\label{eq:div_disc}
\end{eqnarray}
where $\phi_e$ is the value of $\phi$ in the middle of edge $e$.

\subsection{Spatial interpolation}

\begin{figure}
\centering
\includegraphics{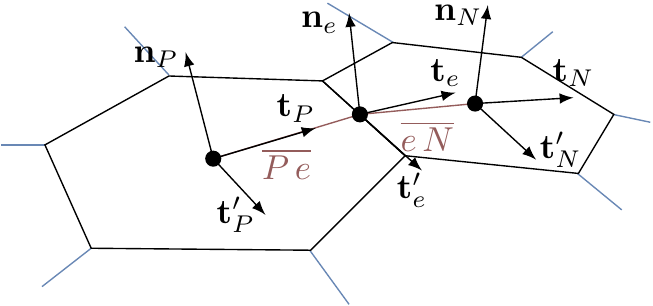}
\caption{Edge-based local coordinate systems and distances for interpolation.}
\label{fig:ca_csys}
\end{figure}

Values, located on the centers of edges $\phi_e$ in Eqs.~\eqref{eq:grad_disc} and \eqref{eq:div_disc}, are interpolated from adjacent cell values. Interpolation of scalars, as flow thickness $h$, can be performed using various schemes. The second-order accurate, unbound, linear, or central interpolation scheme \cite[pp. 76--77]{ferziger2002computational} is given by:
\begin{equation}
h_e = e_x\,h_P + (1-e_x)\,h_N\label{eq:interpolation}
\end{equation}
where $e_x$ is the interpolation factor, calculated as the ratio of distances $\overline{e\,N}$ and $\overline{P\,e\,N} = \overline{P\,e}+\overline{e\,N}$ (Fig.~\ref{fig:ca_csys}):
\begin{equation}
e_x = \dfrac{\overline{e\,N}}{\overline{P\,e\,N}}.
\end{equation}
The central interpolation scheme may cause oscillations, which can be prevented by applying a less accurate first-order scheme. First-order, unconditionally bounded, upwind interpolation \cite[pp. 77--78]{ferziger2002computational} is performed using:
\begin{equation}
e_x = 
\left\{
    \begin{array}{lll}
1 & \mbox{for } &\b{m}_e\bs{\cdot}\b{\o{u}}_e \geq 0 \\ 
0 & \mbox{else}
  \end{array} \right..
\end{equation}
Local blending between linear and upwind interpolation (e.g., hybrid interpolation) is performed using the Gamma interpolation scheme, a normalized variable diagram (NVD) scheme for unstructured meshes \citep{jasak1999gamma}. Diffusivity and stability of the Gamma scheme can be controlled using constant $\beta$, with useful values between $1/10$ and $1/2$, and in which higher values are rather diffusive and lower values are rather unstable.

Vectors cannot be interpolated directly without violating the requirement of a surface-tangential velocity. To address this issue, the interpolation of vectors is performed in local edge-based coordinate systems \citep{tukovic2012moving}, as shown in Fig.~\ref{fig:ca_csys}. Local coordinate systems for cell centers are defined by the surface-normal vectors $\b{n}_P$ and $\b{n}_N$, and unit vectors tangential to lines $\overline{P\,e}$ and $\overline{e\,N}$. The local coordinate system for the edge is defined by the surface-normal vector $\b{n}_e$ and the tangent vector on line $\overline{PeN}$, $\b{t}_e$.
The interpolation of vectors, e.g., velocity, is performed using:
\begin{equation}
\o{\b{u}}_e = \bt{T}^{-1}_e\,\left(e_x\,\bt{T}_P\,\o{\b{u}}_P + (1-e_x)\,\bt{T}_N\,\o{\b{u}}_N\right).
\end{equation}
Herein, $\bt{T}_P$, $\bt{T}_e$, and $\bt{T}_N$ are transformation tensors from the global coordinate system to the local ones. This procedure ensures that interpolations remain surface-tangential. Interpolation factor $e_x$ is calculated as previously, based on the various interpolation schemes.

\subsection{Boundary conditions}

If edge $e$ is placed on border $\partial \Gamma$ of simulation domain $\Gamma$, the interpolation method can not be applied. Instead, a boundary condition has to provide the information required to calculate edge values $\phi_e$. Dirichlet (or fixed value) boundary conditions prescribe the value at the border $\phi(\b{x})=\phi_B \ \forall \b{x} \in \partial\Gamma$, directly yielding the desired edge value:
\begin{equation}
\phi_e = \phi_B.
\end{equation}
Von Neumann (or fixed gradient) boundary conditions, on the other hand, prescribe the gradient at the border. For the simplest case, $\bs{\nabla}\,\phi(\b{x}) = 0\ \forall \b{x} \in \partial\Gamma$, edge values follow as:
\begin{equation}
\phi_e = \phi_P.
\end{equation}
These simple cases are sufficient to provide inflow (Dirichlet boundary conditions for flow thickness and velocity) and outflow (Von Neumann boundary conditions for flow thickness and velocity) boundary conditions for the given shallow flow model \cite[pp. 81--82]{ferziger2002computational}. More complex boundary conditions (e.g., \citep[][pp. 92--96]{jasak1996error}) are not required for our purposes.

\subsection{Temporal discretization}

Discretization of the time derivative at time $t^{n}$ is performed with an implicit, second-order accurate scheme, called a backward scheme \cite[p. 142]{ferziger2002computational}:
\begin{equation}
\left(\dfrac{\partial\,\phi}{\partial t}\right)^n \approx \dfrac{3\,\phi^n-4\,\phi^{n-1}+\,\phi^{n-2}}{2\,\Delta t}.\label{eq:temporal_basis}
\end{equation}
The new, i.e., unknown, time level is referred with index $n$, while old time levels are referred with index $n-1$ and $n-2$, $t^n = t^{n-1} + \Delta t = t^{n-2} + 2\,\Delta t$. This scheme requires the evaluation of the differential equation at only one time step. The differential equation is evaluated at new time level, $t^n$, and therefore the numerical integration is called implicit.

Adaptive time step control can be implemented by considering the Courant number. The Courant number can be calculated as:
\begin{equation}
C = \dfrac{c_e\,\Delta t}{\Delta_e} \leq C_{\r{max}},
\end{equation}
where 
\begin{equation}
c_e = \max\left(\b{m}_e\bs{\cdot}\b{\o{u}}_e\pm\sqrt{\b{g}\bs{\cdot}\b{n}_e\,h_e}\right)
\end{equation}
is the maximum, edge-normal, characteristic wave speed (e.g., \citep[][]{whitham1999linear}) at edge $e$ and $\Delta_e$ is a measure for the cell size, e.g., the distance between centers of adjacent cells, $\overline{P\,e\,N}$. The Courant number is limited to the predefined value $C_{\r{max}}$ to adjust the time step $\Delta t$. Although the applied method is unconditionally stable in terms of time step (the Courant-Friedrichs-Lewy condition, $C_{\r{\max}} = 1$, does not apply), this value constitutes a good indicator for an appropriate time step \cite[pp. 142--151]{ferziger2002computational}. Fixed factors in Eq.~\eqref{eq:temporal_basis} change for adaptive time stepping. It is worth noting that the temporal discretization scheme does not play an important role in the main statement of this paper and can be replaced with any scheme.

\subsection{Initial condition}

The initial condition has to provide $\phi^{n-1}$ on the whole spatial domain $\Gamma$, as required by the time derivative, Eq.~\eqref{eq:temporal_basis}. To avoid the necessity to provide an additional time level $\phi^{n-2}$, the first time step is conducted using the first-order implicit Euler scheme, given by
\begin{equation}
\left(\frac{\partial\,\phi}{\partial t}\right)^n \approx \frac{\phi^n-\phi^{n-1}}{\Delta t}.\label{eq:temporal_basis_reduced}
\end{equation}
For a granular avalanche, a flow thickness $h\neq0$ in the release zone is provided. The initial velocity is usually zero, as well as the flow thickness outside of the release zone.

\subsection{Solution strategy}

The mathematical model forms a system of coupled nonlinear differential equations. The discretization leads to a set of algebraic equations which can be solved numerically. Following the sequential approach, Eq.~\eqref{eq:differential_normal_mom} is solved first for basal pressure $p$ with old solutions of flow thickness $h$ and velocity $\o{\b{u}}$. In a second step, Eq.~\eqref{eq:differential_tang_mom} is solved for velocity $\o{\b{u}}$ with updated values of basal pressure $p$ and old values for flow thickness $h$. Finally, Eq.~\eqref{eq:differential_mass} is solved for flow thickness $h$. The algorithm is iterated until a certain residual is reached. To linearize all nonlinear terms, values of old solutions are introduced. They are marked with an asterisk ($*$) in the following. Terms marked with $n$ are solved implicitly. Convergence is indicated by $\phi^*\approx\phi^n$, fulfilling the nonlinear equations. The fully discretized set of equations reads
\begin{eqnarray}
&&  \dfrac{1}{\rho}\,p_P^{n}\,\b{n}_P\,S_P
+ \left(\b{n}_P\,\,\b{n}_P\right) \bs{\cdot} \xi\,\sum\limits_e{\b{m}_e\bs{\cdot} h^{*}_e\,\o{\b{u}}^{*}_e}\,{\o{\b{u}}^{*}_e}\,L_e\nonumber\\
&&= \left(\b{n}_P\,\b{n}_P\right)\bs{\cdot}h^{*}_P\,\b{g}\,S_P
- \left(\b{n}_P\,\b{n}_P\right) \bs{\cdot} \dfrac{1}{2\,\rho}\,\sum\limits_e\b{m}_e\,h^{*}_e\,p^{*}_e\,L_e \label{eq:fully_discretized_pressure}\\
&&\dfrac{3\,h_P^*\,{\o{\b{u}}_P^n}-4\,h_P^{n-1}\,\o{\b{u}}_P^{n-1}+\,h_P^{n-2}\,\o{\b{u}}_P^{n-2}}{2\,\Delta t}\,S_p
+ \left(\bt{I}-\b{n}_P\,\b{n}_P\right) \bs{\cdot} \xi\sum\limits_e{\b{m}_e\bs{\cdot} h^{*}_e\,\o{\b{u}}^{*}_e}\,{\o{\b{u}}^n_e}\,L_e \nonumber\\
&&= -\dfrac{1}{\rho}\,\mu\,p_P^{n}\,\dfrac{1}{|\o{\b{u}}^{*}_P|+u_0}\,{\o{\b{u}}^n_P}\,S_P
+ \left(\bt{I}-\b{n}_P\,\b{n}_P\right) \bs{\cdot} h^{*}_P\,\b{g}\,S_P \nonumber\\
&&- \left(\bt{I} - \b{n}_P\,\b{n}_P\right) \bs{\cdot} \dfrac{1}{2\,\rho}\,\sum\limits_e\b{m}_e\,h^{*}_e\,p^{n}_e\,L_e ,\label{eq:fully_discretized_momentum}\\
&&\dfrac{3\,{h_P^n}-4\,h_P^{n-1}+\,h_P^{n-2}}{2\,\Delta t}\,S_p
+ \sum\limits_e\b{m}_e\bs{\cdot}{h_e^n}\,\o{\b{u}}^{n}_e \,L_e  = 0.\label{eq:fully_discretized_mass}
\end{eqnarray}
Coupling between cells is represented in the discretized equations by edge values. These values are replaced by cell values according to the interpolation rule. Therefore, the equations contain $\b{\o{u}}^n_N$ and $\b{\o{u}}^n_P$, as well as $h^n_N$ and $h^n_P$. 
Eq.~\eqref{eq:fully_discretized_pressure} is solved explicitly, using old solutions for all fields to obtain updated values for $p_P$. The old solution is either the final solution of an old time step or the last solution of a time step, in which the desired convergence has not yet been achieved.
In case of the momentum Eq.~\eqref{eq:fully_discretized_momentum}, flow thickness $h^{*}$, volume flux \mbox{$\b{m}_e\bs{\cdot} (h^{*}_e\,\o{\b{u}}^{*}_e)\,L_e$} and the magnitude of velocity in friction term $|\o{\b{u}}^{*}|$ are taken from old solutions to obtain a linear system of equations. Other than in the pressure equation, values of adjacent cells $\o{\b{u}}_N^{n}$ are not eliminated. This represents the implicit coupling of adjacent cells, and a solution can only be obtained by solving a linear system of equations, combining the equations for all cells. All scalar coefficients of $\o{\b{u}}_P^n$ and $\o{\b{u}}_N^n$ are summarized to $a_P^u$ and $a_N^u$, respectively. Non-scalar coefficients, e.g., transformation matrices $\bt{T}_P$, $\bt{T}_N$, $\bt{T}_e$, and factor $(\bt{I}-\b{n}_P\,\b{n}_P)$ introduce a coupling between different components of velocity. This is handled explicitly with a deferred correction approach \citep[pp. 122--124]{ferziger2002computational} to keep the system of equations as small as possible. All terms containing neither $\o{\b{u}}_P^n$ nor $\o{\b{u}}_N^n$ are considered as explicit source terms, and are summarized to $\b{b}_P^u$. The source terms include the deferred correction, as well. Eq.~\eqref{eq:fully_discretized_momentum} can then be written as:
\begin{equation}
a_P^u\,\o{\b{u}}_P^n + \sum\limits_N\,a_N^u\,\o{\b{u}}_N^n = \b{b}_P^u,\label{eq:gsys_u}
\end{equation}
and Eq.~\eqref{eq:fully_discretized_mass} similarly as
\begin{equation}
a_P^h\,h_P^n + \sum\limits_N\,a_N^h\,h_N^n = b_P^h.\label{eq:gsys_h}
\end{equation}
Equations~\eqref{eq:gsys_u} and \eqref{eq:gsys_h} can be assembled for each control area $P$, forming a linear system of algebraic equations:\footnote{In the following we apply the Einstein summation convention.}
\begin{eqnarray}
A_{ij}^u\,\b{\o{u}}^n_j = \b{b}_i^u,\\
A_{ij}^h\,h^n_j = b_i^h,
\end{eqnarray}
where $A_{ij}^u$ and $A_{ij}^h$ are coefficients of sparse matrices (diagonal coefficients $a_P^u$, $a_P^h$ and off diagonal coefficients $a_N^u$, $a_N^h$). $\b{\o{u}}^n_j$ and $h^n_j$ are the depth-averaged velocity and the flow thickness, respectivly, of the $j^{\r{th}}$ computational point at the new time level $t^n$. The algebraic systems are solved using an iterative solution procedure, suitable for sparse matrices. We  achieved the best stability with a preconditioned (diagonal incomplete-LU) bi-conjugate gradient method (DILU-PBiCG, \cite{opencfd2009user}). The residual $r^{\phi}$, used to check for convergence, is calculated as:
\begin{equation}
r^{\phi} = \dfrac{\sum\limits_i \left|b_i^\phi-A_{ij}^\phi\,\phi^n_j\right|}{f_n^\phi},\label{eq:residual}
\end{equation}
where $\phi$ is either velocity $\b{\o{u}}$ or flow thickness $h$. $f_{\r{n}}^\phi$ is the normalization factor, calculated as:
\begin{equation}
f_{\r{n}}^{\phi} = \sum\limits_i \left|A_{ij}^\phi\,\phi_j-A_{ij}^\phi\,\langle \phi^n \rangle\right|+\left|b_i^\phi-A_{ij}^\phi\,\langle \phi^n \rangle\right|,
\end{equation}
and 
\begin{equation}
\langle \phi^n \rangle = \dfrac{1}{J}\sum\limits_{j=0}^J\phi^n_j
\end{equation}
is the average of $\phi^n_j$ \citep{opencfd2009user}. It is worth noting that the residual will be a vector if $\phi$ is a vector, as for velocity. 
The residual is utilized to check the convergence of the linear system (final residual). Moreover, it is used to check the convergence of the coupled system, by calculating it before values of $\phi_j^n$ are updated, e.g., by applying $\phi^*_j$ instead of $\phi^n_j$ in Eq.~\eqref{eq:residual} (initial residual). To achieve convergence, matrices are under-relaxed \cite[pp. 129--130]{ferziger2002computational}.

Dry areas, where $h=0$, are handled on the matrix level. In the case of $h=0$, the respective diagonal element of velocity equation matrix $a_P^{u}$ is less than or equal to zero, and the matrix is singular or negative-definite. On these cells, we aim to obtain $\b{\o{u}} = \b{0}$, which can be forced by setting $a_P^{u} = 1$, $a_N^{u} = 0$, and $\b{b}_P^{u} = \b{0}$, leading to a positive-definite matrix, as well. Alternatively, flow thickness can be bound to a certain value $h_{\r{min}} > 0$, effectively eliminating any dry areas. Both approaches have been evaluated, and equal results have been obtained. The first method is preferred, since the second approach leads to numerical artifacts (e.g., $\b{\o{u}}  \neq \b{0}$) in areas that are supposed to be dry.

\section{Implementation}
\label{sec:implementation}

The finite area method, as described in section~\ref{sec:fam}, was first introduced by Tukovi{\'c} and Jasak \cite{tukovic2005metoda, tukovic2012moving} based on the CFD toolkit OpenFOAM \citep{weller1998tensorial, opencfd2009pro, opencfd2009user} to solve surface transport equations. Its original implementation is currently part of the branch foam-extend.

OpenFOAM is written in C++, and utilizes object-oriented techniques and operator overloading to form a top-level syntax similar to the tensor notation of partial differential equations. A major component of OpenFOAM, the finite volume (FV) library, solves three-dimensional partial differential equations (usually Navier-Stokes equations) on unstructured, polyhedral meshes. The finite area (FA) library solves quasi-two-dimensional partial differential equations (e.g., shallow flow equations) on unstructured polygonal meshes. In practice, the FA mesh is the boundary of the FV mesh \citep[][]{tukovic2012moving}. This is especially advantageous for a coupling of surface transport equations with three-dimensional ambient flows. Moreover, the mesh can be created and manipulated with a wide range of tools.

The finite area library provides both implicit and explicit versions of the divergence and gradient operators, and their respective projections.
Implicit operators, contributing to factors $a_N$ and $a_P$ (partially also to $b_P$ due to deferred correction and the source part of the transistent term), are part of the namespace \texttt{fam} (namespaces are used in C++ to distinguish different kinds of functions). Explicit operators are solely contributing to the source term $b_P$ and are part of the namespace \texttt{fac}.

The top level solver code mimics the structure of Eqs.~\eqref{eq:differential_normal_mom}, \eqref{eq:differential_tang_mom}, and \eqref{eq:differential_mass}. Pressure is calculated explicitly, and the respective code reads:
\begin{verbatim}
    pbn = rho*                          //basal pressure
    (
      - xi*fac::ndiv(F2s, Us)           //centrifugal forces
      + ((g*h)&n)*n                     //gravity
      - fac::ngrad((pbn&n)*h/(2.*rho))  //lateral pressure
    );
\end{verbatim}
The assembly of the momentum equation reads:
\begin{verbatim}
        fam::ddt(h, Us)                 //transistent term
      + xi*fam::div(F2s, Us)            //convective term
      + fam::Sp(nu, Us)                 //basal friction
     ==
        g*h - ((g*h)&n)*n               //gravity
      - fac::grad((pbn&n)*h/(2.*rho))   //lateral pressure    
\end{verbatim}
Finally, the assembly of the continuity equation is:
\begin{verbatim}
        fam::ddt(h)                     //transistent term
      + fam::div(Fs, h)                 //convection
\end{verbatim}
\texttt{Fs} and \texttt{F2s} represent explicitly calculated edge fluxes, and \texttt{nu} represents an effective viscosity, calculated by a runtime selectable basal friction model. For the presented rheology, it is calculated as:
\begin{equation}
\nu = \dfrac{1}{\rho}\mu\left(I_{\r{b}}\right)\,p_{\r{b}}\,\dfrac{1}{|\o{\b{u}}|+u_0}.
\end{equation}
The assembly of the momentum and continuity equations is followed by a call of the matrix solver routine. These parts are embedded in a loop iterating until the fields do not change significantly.\\
The link between implementation and the mathematical model allows a simple extension of the basic solver. Considering complex physics in granular flows, debris flows and avalanches, this is an important feature allowing rapid development and testing of novel hypotheses.

\section{Affinities to former approaches}
\label{sec:affinity}

\begin{figure}
\centering
\includegraphics{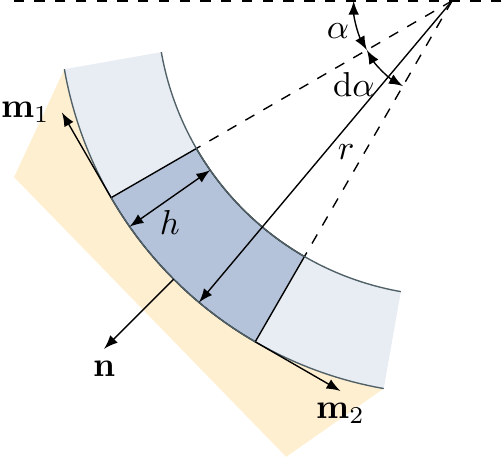}
\caption{Quasi one-dimensional free surface flow on a constantly curved surface.}
\label{fig:example1d}
\end{figure}

Before showing results of the numerical implementation, we aim to establish a link to the extended Savage-Hutter model \citep{greve1994unconfined}, considering a simply curved, quasi one-dimensional slope in steady state. 
This demonstrates the equivalence of curvature-free expressions from this work and the curvature terms of classic approches.
Flow thickness $h$, magnitude of depth-averaged velocity $|\b{\o{u}}|$, and curvature $\kappa$ are assumed to be constant. Following the traditional approach of Greve et al. \cite{greve1994unconfined}, the surface pressure due to centrifugal forces is:
\begin{equation}
p_{\r{c}} = \kappa\,\rho\,h\,|\b{\o{u}}|^2,\label{eq:sh_pressure}
\end{equation}
and following the approach represented in this paper:
\begin{equation}
p_{\r{c}} = -\dfrac{1}{S_{\r{b}}}\rho\sum\limits_e \b{m}_e\bs{\cdot} h_e\,\b{\o{u}}_e\,\b{\o{u}}_e\,L_e\bs{\cdot}\b{n}.\label{eq:my_pressure}
\end{equation}
Fig.~\ref{fig:example1d} shows a small section between $\alpha$ and $\alpha+\r{d}\alpha$ of the slope. Following this figure, normal and bi-normal vectors can be expressed as:
\begin{equation}
\b{m}_1 = \left(
\begin{array}{c}
-\sin(\alpha)\\
\cos(\alpha)
\end{array}\right),
\end{equation}
\begin{equation}
\b{m}_2 = \left(
\begin{array}{c}
\sin\left(\alpha+\r{d}\alpha\right)\\-\cos\left(\alpha+\r{d}\alpha\right)
\end{array}\right),
\end{equation}
\begin{equation}
\b{n}  = \left(
\begin{array}{c}-\cos\left(\alpha+\dfrac{\r{d}\alpha}{2}\right)\\
-\sin\left(\alpha+\dfrac{\r{d}\alpha}{2}\right)\end{array}\right),
\end{equation}
and surface area $S_{\r{b}}$ as:
\begin{equation}
S_{\r{b}} = 2\,r\,\sin\left(\dfrac{\r{d}\alpha}{2}\right).
\end{equation}
The flow thickness is constant and velocities can be expressed as
\begin{equation}
\b{\o{u}}_e = \pm|\b{\o{u}}|\,\b{m}_e,
\end{equation}
leading to:
\begin{equation}
p_{\r{c}} = -\rho\,h\,{|\b{\o{u}}|}^{2}\,\dfrac{1}{S_{\r{b}}}\sum\limits_e \b{m}_e\,L_e\bs{\cdot}\b{n}.\label{eq:kappa_hint}
\end{equation}
Introducing values for $\b{m}_e$, $\b{n}$ and $S_{\r{b}}$, and simplifying the relation leads to:
\begin{equation}
p_{\r{c}} = \rho\,h\,|\b{\o{u}}|^2\, \dfrac{2\,\sin\left(\dfrac{\r{d}\alpha}{2}\right)}{2\,r\,\sin\left(\dfrac{\r{d}\alpha}{2}\right)} = \dfrac{1}{r}\rho\,h\,|\b{\o{u}}|^2,\label{eq:p_c}
\end{equation}
matching relation~\eqref{eq:sh_pressure}.
Basal pressure, associated with curvature in combination with lateral pressure, can be calculated in a similar manner, approximately yielding:
\begin{equation}
p_{\r{lp}} \approx \dfrac{1}{2}\,\rho\,\dfrac{h^2}{r^2}\,|\b{\o{u}}|^2 + \dfrac{1}{2}\rho\,g_{\r{n}}\dfrac{h^2}{r}.\label{eq:p_fb}
\end{equation}
This term is approximately by a factor of $\frac{1}{2}\frac{h}{r}$ smaller than other contributions to pressure. Considering that curvature radius $r$ is usually much larger than flow thickness $h$, this term can be safely neglected as in other approaches. However, to achieve a three-dimensional momentum equilibrium, this term is included in our simulations.

\section{Results and discussion}
\label{sec:results}

The implementation of the finite area method has been used and verified for several problems \citep{tukovic2012moving, marschall2014validation, dieter2015direct, dieter2015numerical, cardiff2016block}, as well as the related finite volume method (e.g., \citep[][]{weller1998tensorial}) and the interpolation scheme \citep{jasak1999gamma}. In this work, we will present four examples with increasing complexity, focusing on different aspects. Some of the results are compared to analytical solutions for quantitative verification. The convergence order is demonstrated, as well.

The first example presents variations of the popular dam break case, including a comparison to analytical solutions.
The second example highlights results for basal pressure, including a comparison to the solution of Greve et al.\ \cite{greve1994unconfined} (see section~\ref{sec:affinity}).
The third example presents a granular avalanche on a simply curved, structured mesh, that is well-suited for a refinement study, demonstrating the convergence order of the method. 
The fourth example shows the handling of complex topography and unstructured meshes.

\subsection{Dam break}

The dam break is the simplest transitory solution for the Shallow Water Equations, and is widely utilized as a benchmark case to demonstrate various features \cite{delestre2013swashes}. In this section, three variations of the dam break case are simulated, showing the capability to represent shocks, the transition between wet ($h>0$) and dry ($h=0$) domains, and source terms (e.g., friction). Analytical solutions for the three variations have been found by Stoker \cite{stoker1957water} , Ritter \cite{ritter1892fortpflanzung}, and Dressler \cite{dressler1952hydraulic}. The analytical solutions are collected and provided by Delestre et al. \cite{delestre2013swashes}.

The dam break is a one-dimensional problem on a flat surface. A fluid reservoir with initial flow thickness $h_l$ spans over half the simulation domain ($x \leq x_0$) and is suddenly released at $t=0$. The flow thickness in the other half ($x > x_0$) $h_r$ is either $0\,\r{m}$ (dry case) or smaller than $h_l$ (wet case). The initial velocity is $0$. 
Herein, the simulation domain spans over $\left[0\,\mathrm{m}, 10\,\mathrm{m}\right]$, the mesh consists of $2\,000$ cells and the dam is located at $x_0= 5\,\r{m}$. We chose $h_l = 0.5\,\r{m}$, $h_r =0.1\,\r{m}$ for the wet case and $h_r = 0$ for the dry case in all simulations. Dry areas have been handled explicitly on the matrix level.

\subsubsection{Stoker's solution - dam break on a wet domain without friction}

\begin{figure}
\includegraphics[scale=0.75, trim=0cm 0cm 2cm 0cm, clip=true]{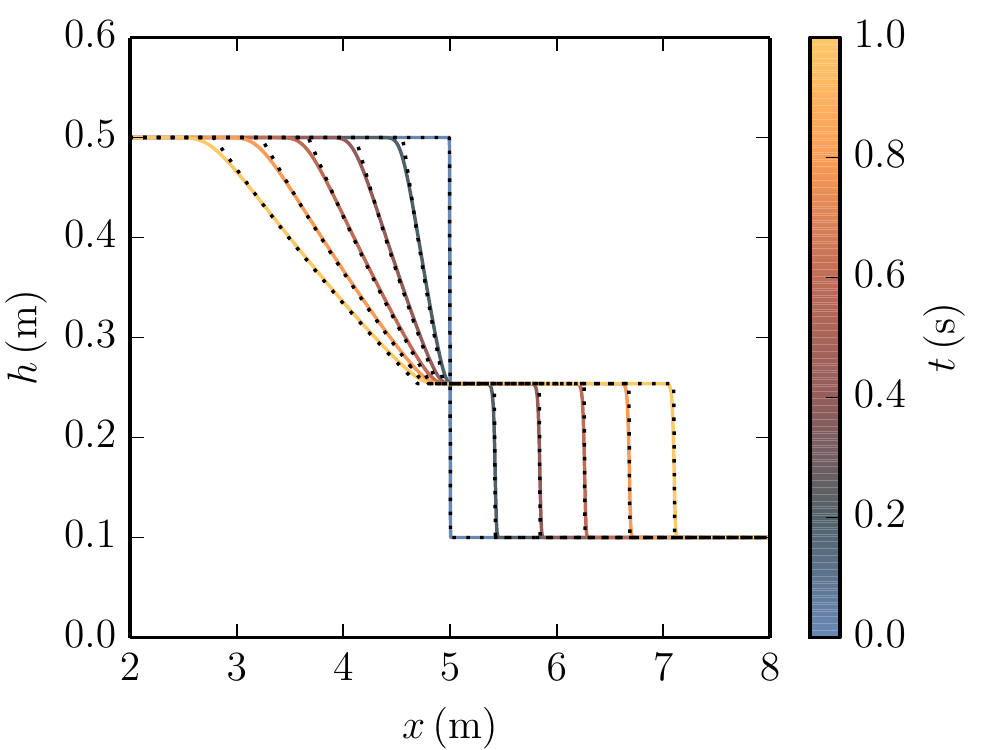}
\hfill
\includegraphics[scale=0.75]{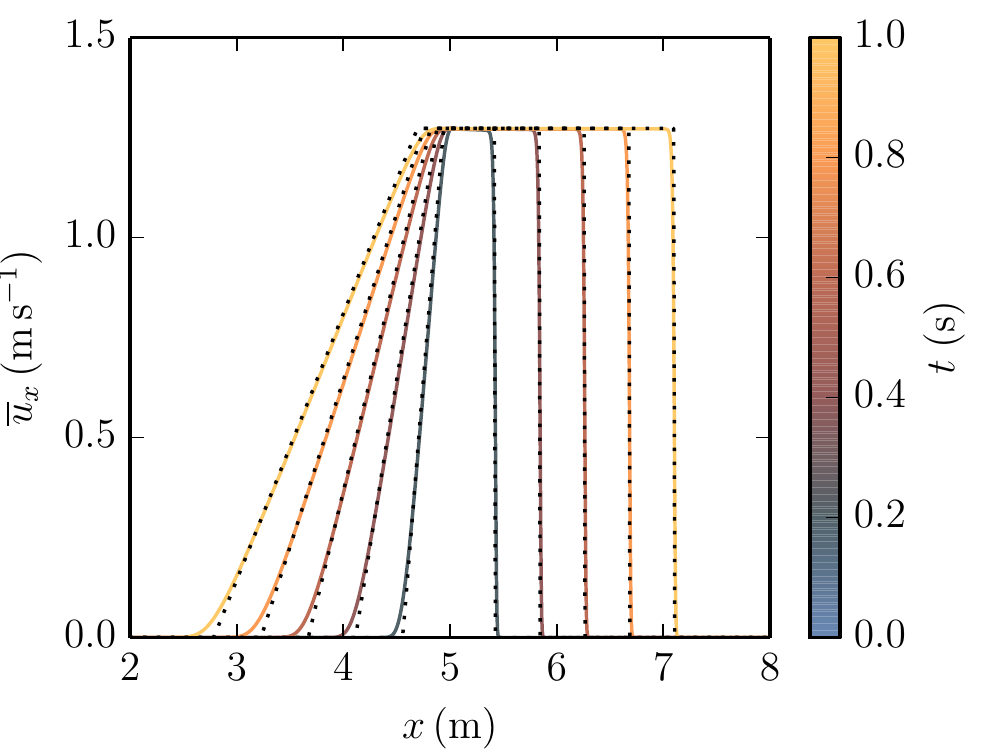}
\caption{Dam break on a wet domain, upwind scheme. Numerical results (colored) for flow thickness (left) and depth-averaged velocity (right) are compared to Stoker's analytical solution (black, dotted).}
\label{fig:stokers_upwind}
\end{figure}

\begin{figure}
\includegraphics[scale=0.75, trim=0cm 0cm 2cm 0cm, clip=true]{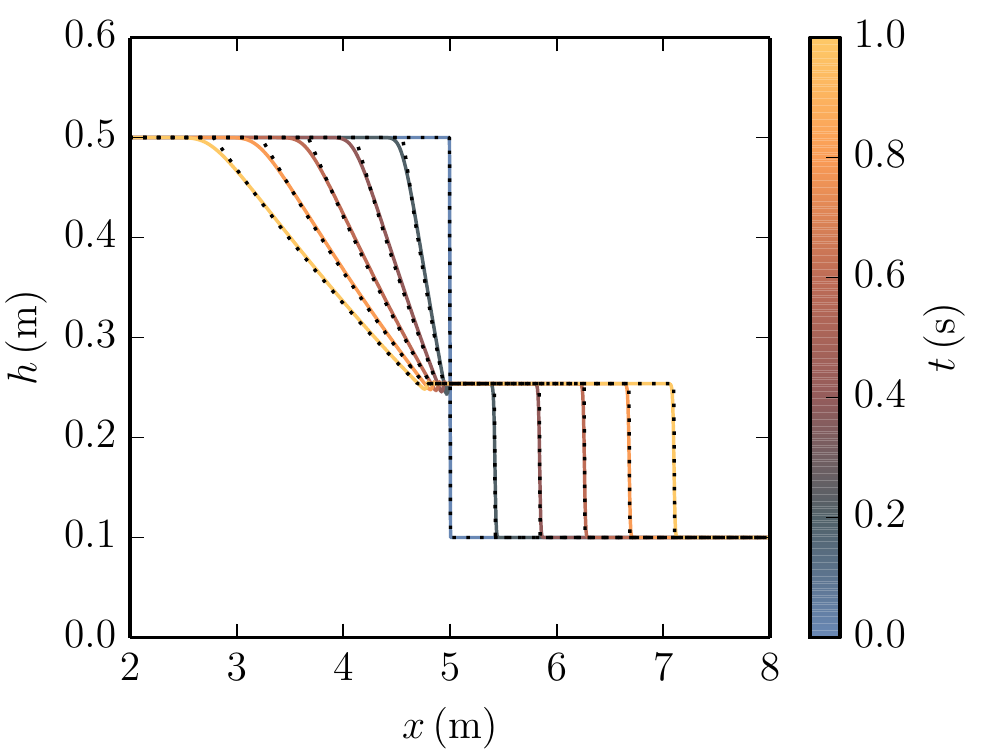}
\hfill
\includegraphics[scale=0.75]{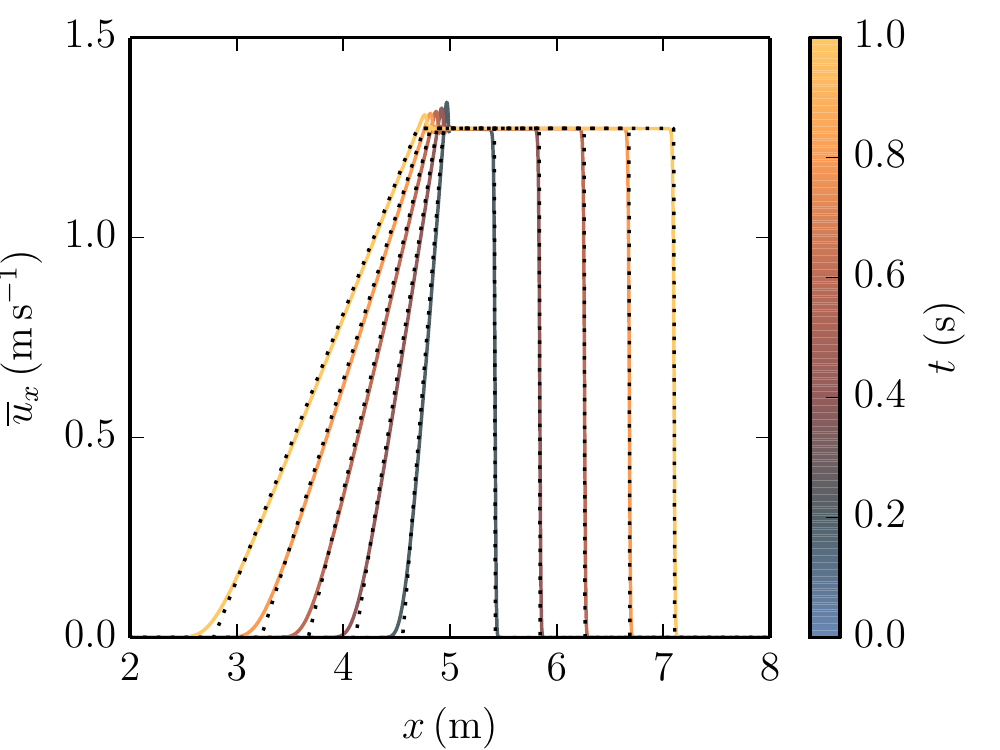}\\
\caption{Dam break on a wet domain, Gamma scheme. Numerical results (colored) for flow thickness (left) and depth-averaged velocity (right) are compared to Stoker's analytical solution (black, dotted).}
\label{fig:stokers_gamma}
\end{figure}

The solution of Stoker \cite{stoker1957water} describes a dam break on a wet domain without friction. This solution is similar to the shock tube \citep{sod1978survey}, known from compressible gas dynamics. This case is used to demonstrate the handling of shocks. Results of the numerical solution are presented in Fig.~\ref{fig:stokers_upwind} for the upwind scheme and Fig.~\ref{fig:stokers_gamma} for the Gamma scheme. The Gamma scheme exhibits some small oscillations around $x=5\,\mathrm{m}$, but a sharper shock wave in comparison to the upwind scheme.


\subsubsection{Ritter's solution - dam break on a dry domain without friction}

\begin{figure}
\includegraphics[scale=0.75, trim=0cm 0cm 2cm 0cm, clip=true]{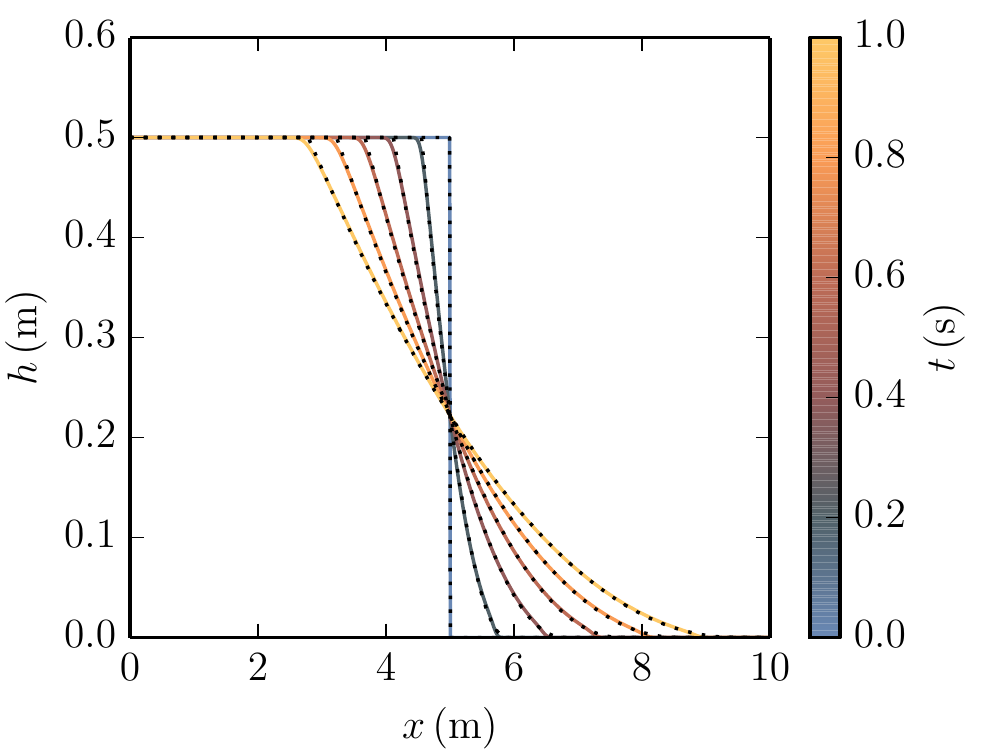}
\hfill
\includegraphics[scale=0.75]{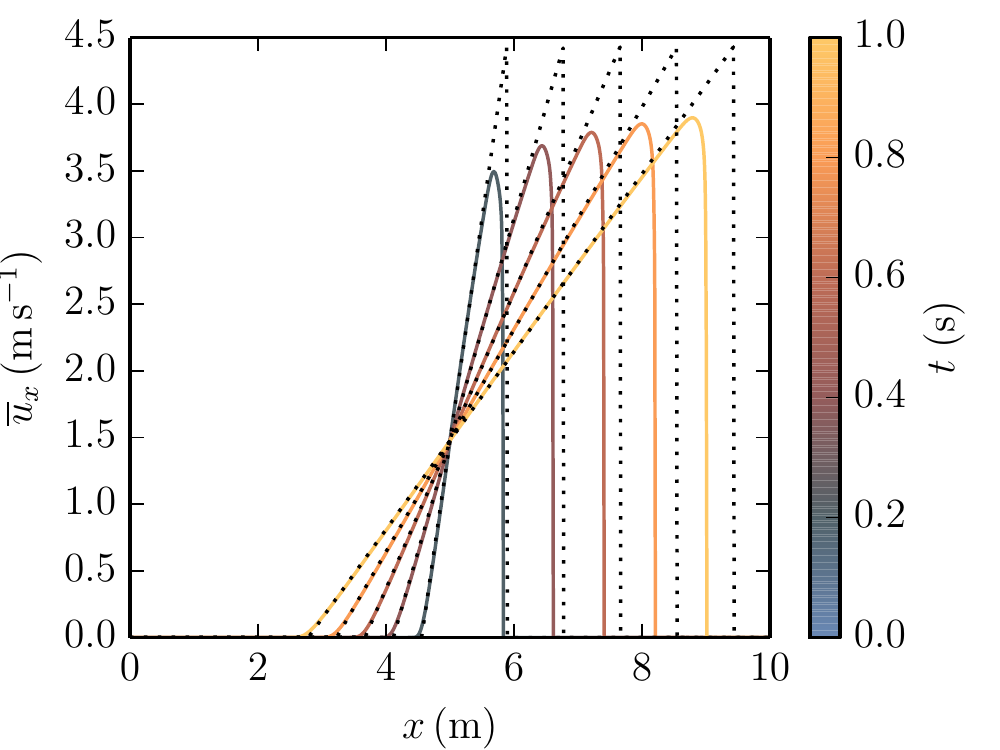}
\caption{Dam break on a dry domain without friction. Numerical results (colored) for flow thickness (left) and depth-averaged velocity (right) are compared to Ritter's analytical solution (black, dotted).}
\label{fig:ritter}
\end{figure}

The solution of Ritter \cite{ritter1892fortpflanzung} describes a dam break on an initially dry domain without friction. Results are shown in Fig.~\ref{fig:ritter} for the Gamma scheme. The solution for velocity at the front cannot be reproduced accurately by the numerical scheme. However, considering the sharp peak, this behavior has to be expected, and refinement of the mesh will reduce the error. On the other hand, the result for flow thickness matches the analytical solution sufficiently.

\subsubsection{Dressler's solution - dam break on a dry domain with friction}

\begin{figure}
\includegraphics[scale=0.75, trim=0cm 0cm 2cm 0cm, clip=true]{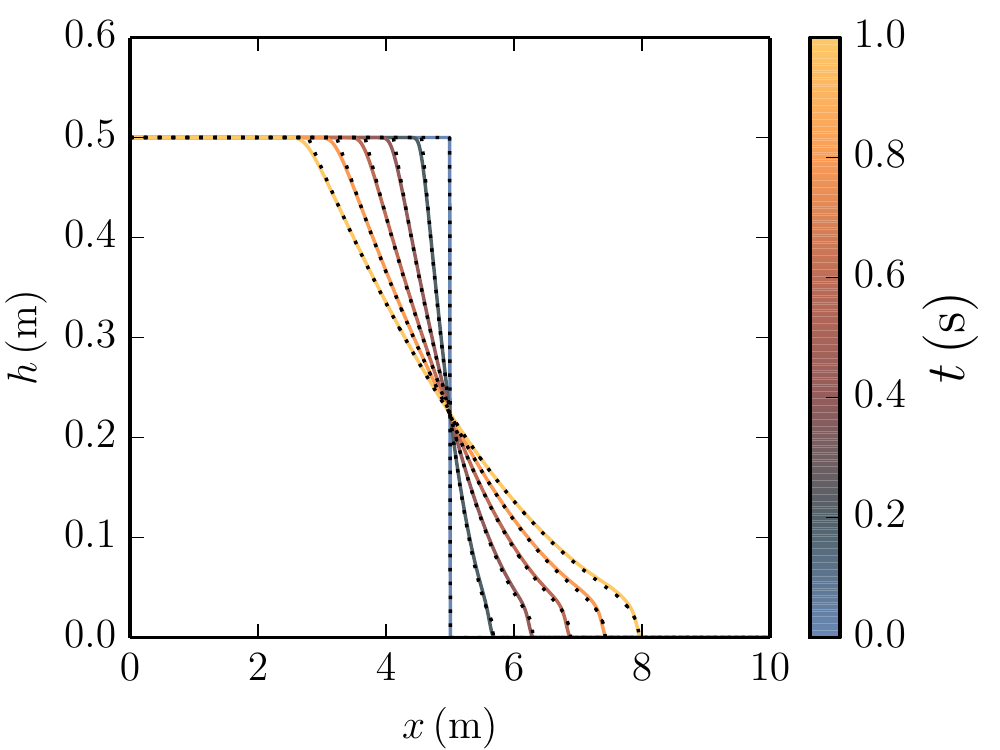}
\hfill
\includegraphics[scale=0.75]{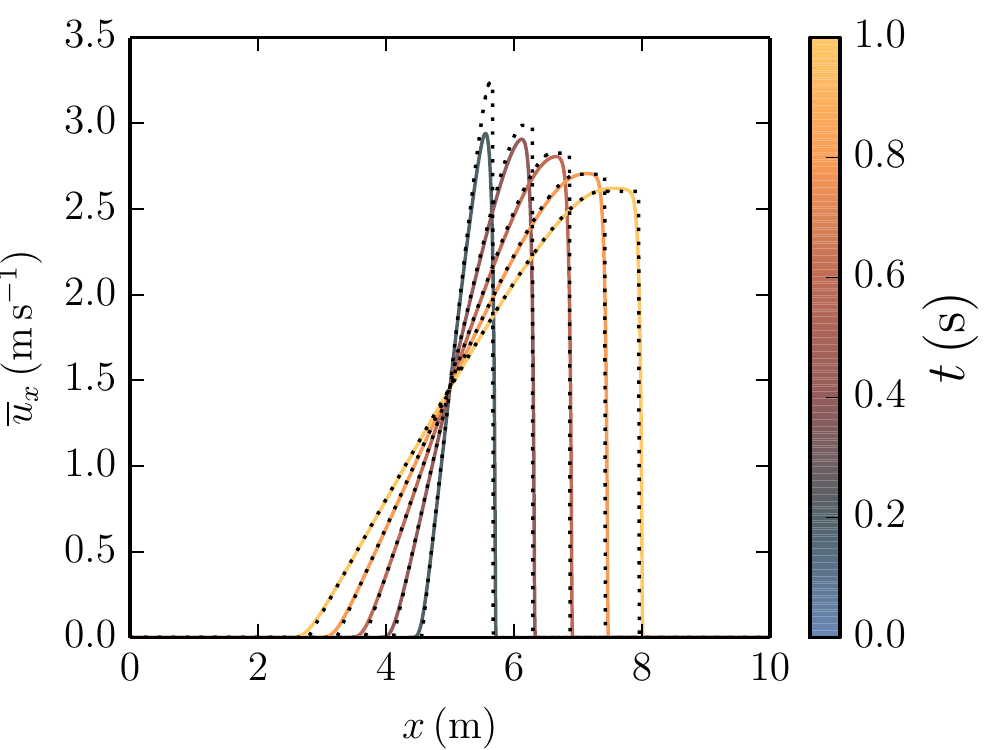}
\caption{Dressler's solution for a dam break case on a dry domain with friction. Numerical results (colored) for flow thickness (left) and depth-averaged velocity (right) are compared to analytical results (black, dotted).}
\label{fig:dressler}
\end{figure}

Finally, the solution of Dressler \cite{dressler1952hydraulic,whitham1955effects} includes the effect of friction following the Ch\'{e}zy friction law, given as:
\begin{equation}
\bs{\tau}_{\r{b}} = \dfrac{\rho\,g}{C^2}|\b{\o{u}}|\,\b{\o{u}}.
\end{equation}
We chose $C = 40\,\r{m^{1/2}\,s^{-1}}$ for the simulation shown in Fig.~\ref{fig:dressler}. The front velocity is clearly damped by the basal friction, allowing the numerical scheme to capture velocity better than in the previous example. It is worth noting that the implementation of Delestre et al. \cite{delestre2013swashes}, which is also used here, contains the semi-analytical modification of Valiani et al. \cite{valiani2004case}.

\subsection{Simply curved sloped}
\label{sec:res_simply}

\begin{figure}
\centering
\includegraphics[width=0.5\columnwidth]{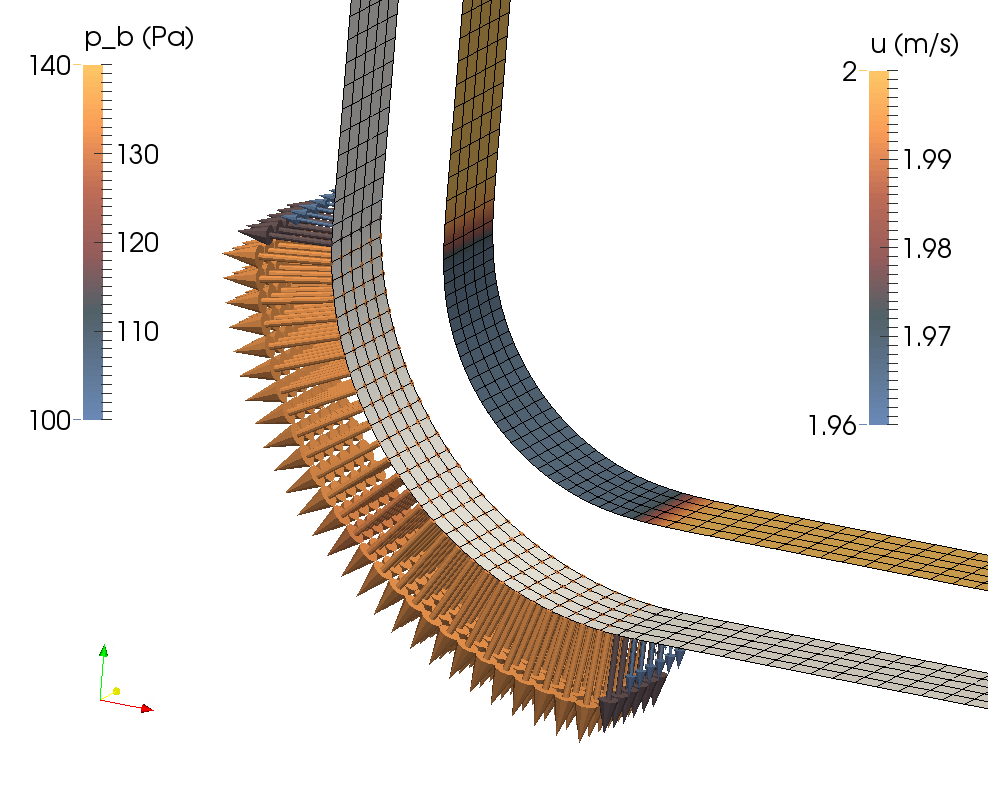}
\caption{Simulation of a quasi one-dimensional free surface flow on a curved surface. The topography surface is the lower one. The free surface is indicated above, colored according to velocity $|\b{\o{u}}|$ (height has been scaled with a factor of $10$). Arrows show the basal pressure, colored accordingly.}
\label{fig:testtransport}
\end{figure}

Here, we aim to present a comparison of numerical results with the analytical results derived in section~\ref{sec:affinity}, regarding basal pressure (Eqs.~\eqref{eq:p_c} and \eqref{eq:p_fb}). Surface-tangential transport equations have been addressed with the popular dam break cases, and this section focuses on the surface-normal equations. Gravity and friction have been set to zero for this example, as they would disturb constant velocity and thickness fields, and therefore complicate the example. The simulation set up, including results, is shown in Fig.~\ref{fig:testtransport}. The fluid enters the curved simulation domain vertically at the upper-left-side (inlet) and leaves the simulation domain at the lower-right-side (outlet). Curvature in this example is $\kappa = \frac{1}{3}\,\r{m^{-1}}$, velocity at the inlet is $2\,\r{m\,s^{-1}}$, flow thickness at the inlet is $0.1\,\r{m}$, and density is $1000\,\r{kg\,m^{-3}}$. When the flow reaches the curved segment, it is slightly slowed down to $1.97\,\r{m\,s^{-1}}$, and the flow thickness increases accordingly to $0.102\,\r{m}$. The basal pressure due to centrifugal forces in the curved segment follows as:
\begin{equation}
p_{\r{c}} = \kappa\,\rho\,h\,|\b{\o{u}}|^2 = 131.1\,\r{N\,m^{-2}},
\end{equation}
the basal pressure due to the feedback of lateral pressure is: 
\begin{equation}
p_{\r{l}} \approx \dfrac{1}{2}\,\kappa^2\,\rho\,h^2\,|\b{\o{u}}|^2=2.2\,\r{N\,m^{-2}},
\end{equation}
effectively yielding a basal pressure of $p_{\r{b}} \approx 133.3\,\r{N\,m^{-2}}$, matching the result of the numerical routine, $p_{\r{b}} = 133.5\,\r{N\,m^{-2}}$ (compare Fig.~\ref{fig:testtransport}). The depth-integrated pressure jump between straight and curved segment follows as $\frac{1}{2}\,p_{\r{b}}\,h = 6.82\,\r{N\,m^{-1}}$, leading to reduced velocity and increased thickness in the curved segment. The pressure jump in this example is small, and taking it into account seems unnecessary. However, the friction associated with pressure makes its accurate description essential, as demonstrated in the next examples.

\subsection{Granular avalanche on a simply curved slope}
\label{sec:res_simply_avalanche}

\begin{figure}
\centering
\includegraphics[width=\columnwidth]{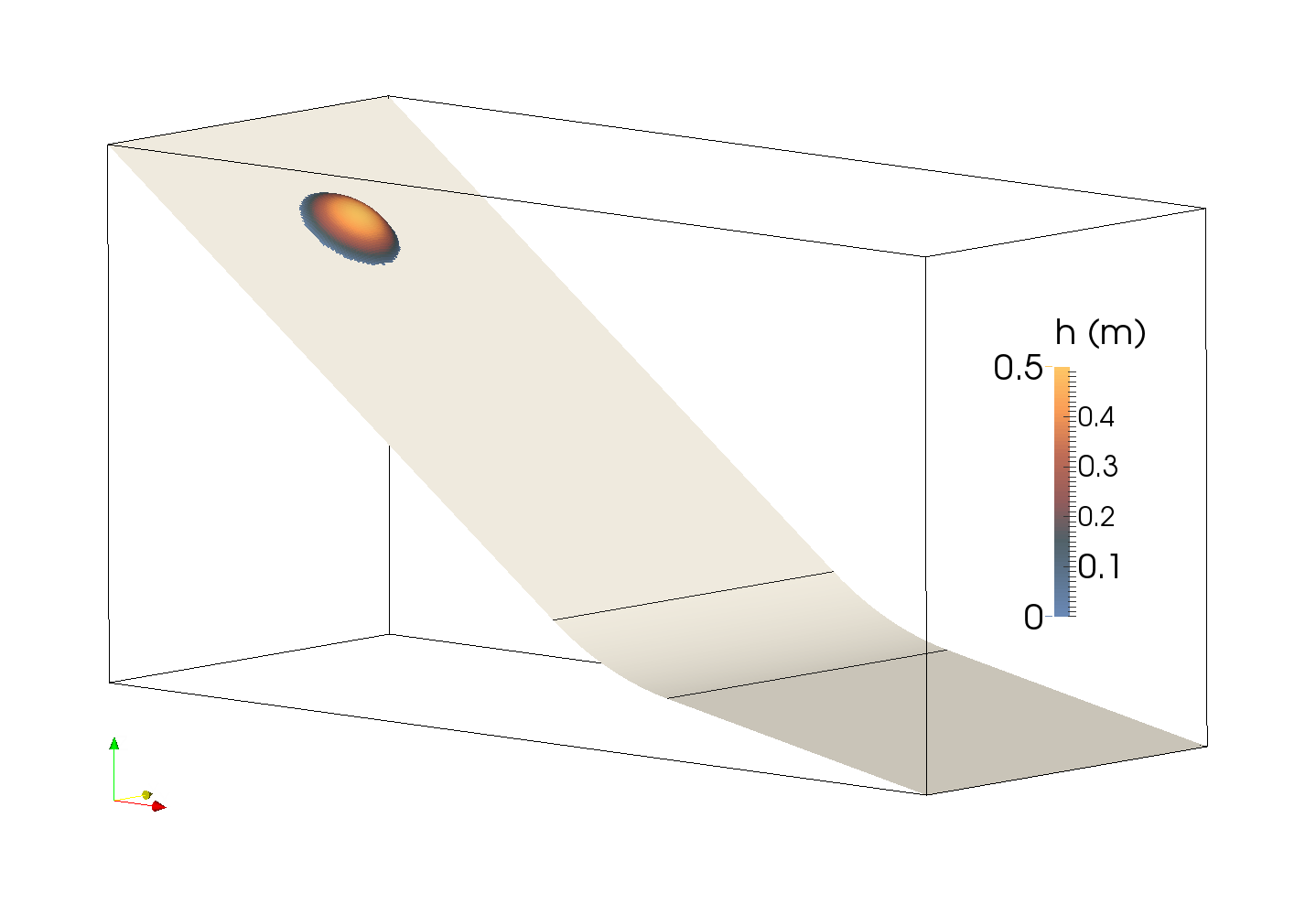}
\caption{Slope geometry and initial condition of the simply curved slope. The color marks the initial flow thickness, and the two horizontal lines mark the transition zone between $\loc{x} = 17.5\,\r{m}$ and $\loc{x} = 21.5\,\r{m}$. }
\label{fig:convergence_initial}
\end{figure}

Here, we show the simulation of a granular avalanche on an inclined slope with a curved transition between two constant inclinations. Similar examples have been widely used for demonstration purposes in the granular flow community \citep{pudasaini2007avalanche}. The inclination angle $\zeta$ is usually described as a function of the local $\loc{x}$-axis. We chose:
\begin{equation}
\zeta = 
\left\{
\begin{array}{ll}
35^\circ &\mbox{for } \loc{x} < 17.5\,\r{m}\\ 
35^\circ-6.25^\circ\,\left(\loc{x}-17.5\,\r{m}\right) &\mbox{for } 17.5\,\r{m} \leq \loc{x} \leq 21.5\,\r{m} \\
10^\circ &\mbox{for } \loc{x} > 21.5\,\r{m}\\ 
\end{array}\right..
\end{equation}
The three-dimensional surface can be reconstructed by integrating trigonometric functions of the inclination, $\r{d}x = \cos(\zeta)\,\r{d}\loc{x}$, $\r{d}y = -\sin(\zeta)\,\r{d}\loc{x}$. Usually, a horizontal ($\zeta = 0^\circ$) runout area is chosen. However, the slight inclination of the runout area of $10^\circ$ increases the sensitivity to physical and numerical parameters, making the example more suitable for a convergence analysis. The essential behavior remains the same, and the characteristics of granular flows are clearly visible in the results. The initial condition is formed by a resting pile ($\b{\o{u}} = \b{0}$) of granular material in the form of a spherical cap with center $\loc{\b{x}}_{\r{c}} = (4, 0, -1.5)^\r{T}$ and radius $r_{\r{c}} = 2\,\r{m}$, leading to a maximum thickness of $0.5\,\r{m}$. Geometry and initial conditions are shown in Fig.~\ref{fig:convergence_initial}.
\noindent
We applied two interpolation schemes to the calculation of the convective term, the upwind scheme, and the Gamma scheme. Interpolations aside convective terms are performed by a linear interpolation. The mesh consists of equal, square cells. The mesh has been gradually refined, down to a minimal size of $\Delta = 0.056\,\r{m}$ ($94\,874$ cells). Numerical and material parameters are listed in Table~\ref{tab:params}.

\begin{table}
\centering
\caption{Numerical and material parameters.}
\label{tab:params}
\begin{tabular}{llc}
parameter & description & value\\
\hline
\multicolumn{3}{l}{\textbf{Physical parameters}}\\
\hline
$g_{\r{z}}$ & gravitational acceleration & $9.81\,\r{m\,s^{-2}}$\\
$\xi$ & shape factor & $1$\\
$\rho$ & flow density & $1500\,\r{kg\,m^{-3}}$\\
$\rho_{\r{p}}$ & particle density & $2500\,\r{kg\,m^{-3}}$\\
$d$ & particle diameter & $0.005\,\r{m}$\\
$\mu_{\r{s}}$ & friction value slow & $0.38$\\
$\mu_{2}$ & friction value fast & $0.65$\\
$I_0$ & reference inertial number & $0.3$\\
\hline
\multicolumn{3}{l}{\textbf{Numerical parameter}}\\
\hline
$u_0$ & threshold viscosity & $10^{-4}\,\r{m\,s^{-1}}$\\
$h_{\min}$ & min.\ flow thickness & $10^{-6}\,\r{m}$\\ 
$\alpha_{\r{u}}$ & relaxation $\b{\o{u}}$ eq. & $0.5$\\
$\alpha_{\r{h}}$ & relaxation $h$ eq. & $0.5$\\
$r_{\r{u},\max}$ & TOL initial residual $\b{\o{u}}$ eq. & $10^{-6}$\\
$r_{\r{h},\max}$ & TOL initial residual $h$ eq. & $10^{-6}$\\
$n_{\max}$ & max.\ iterations per step & $50$\\
$C_{\r{max}}$ & max.\ Courant number & $1$\\
$\beta$ & Gamma scheme const. & $0.5$\\
\hline
\end{tabular}
\end{table}

\begin{figure}
\centering
\includegraphics[scale=0.75]{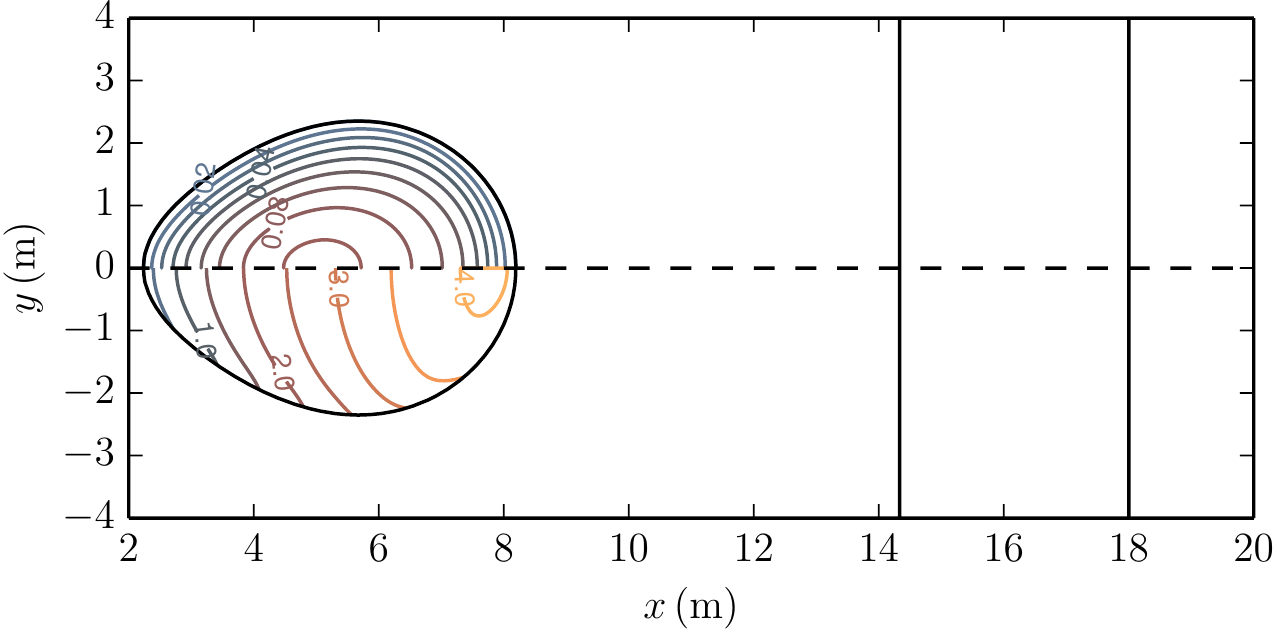} \includegraphics[scale=0.75]{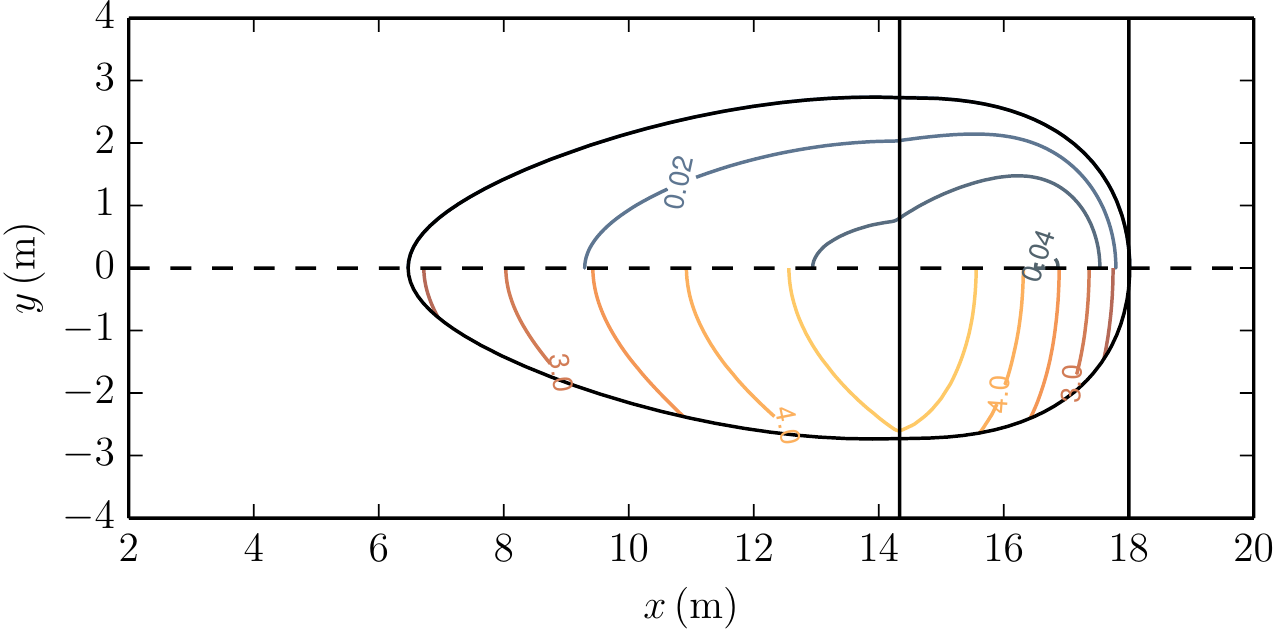} \includegraphics[scale=0.75]{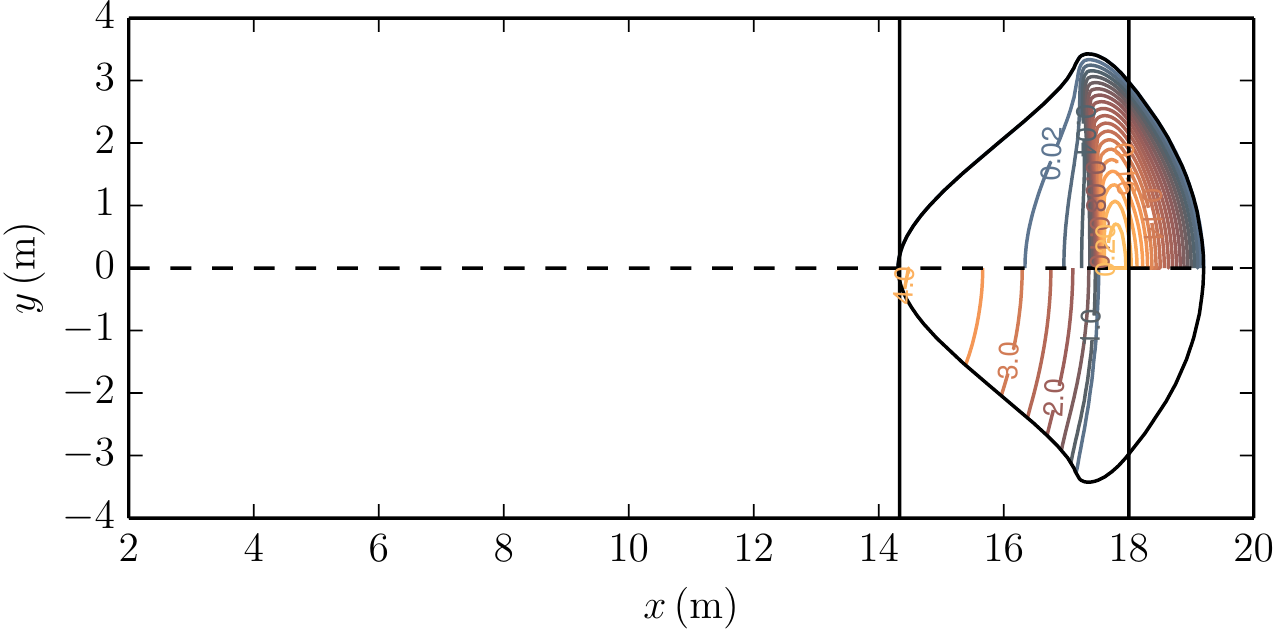} \includegraphics[scale=0.75]{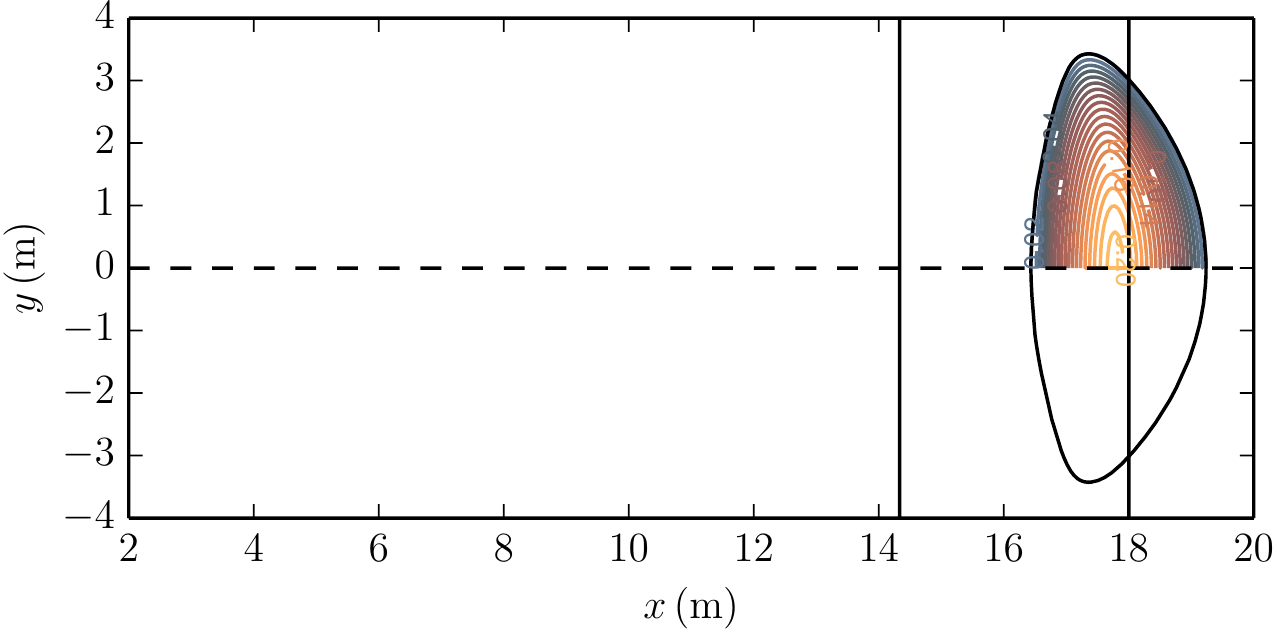}
\caption{Top view of the slope for times $t=2\,\r{s}$, $t=4\,\r{s}$, $t=6\,\r{s}$, and the final position at $t=25\,\r{s}$ for the upwind scheme and a mesh size of $0.056\,\r{m}$. Contour lines indicate the flow thickness in $0.01\,\r{m}$ steps, up to $0.2\,\r{m}$ (top half) and the depth-averaged velocity in $0.5\,\r{m\,s^{-1}}$ steps, up to $4.5\,\r{m\,s^{-1}}$ (bottom half). Areas in which the flow thickness is lower than $0.01\,\r{m}$ are not shown. For orientation, the central flow path, matching the x-axis, and the transition zone are shown.}
\label{fig:simple_slope_run}
\end{figure}

Fig.~\ref{fig:simple_slope_run} shows the simulation with the finest mesh at different time steps for the upwind scheme. Simulations with the Gamma scheme appear less smooth due to the lower diffusion. The shock in the transition zone is visible at time $t=6\,\r{s}$. Results qualitatively match the former solution (e.g., \cite[][]{wang2004savage}). Differences can be attributed to rheology (shear rate dependence and earth pressure coefficients) and the geometry of the slope (inclination of the runout zone).

\begin{figure}
\centering
\includegraphics[scale=0.67]{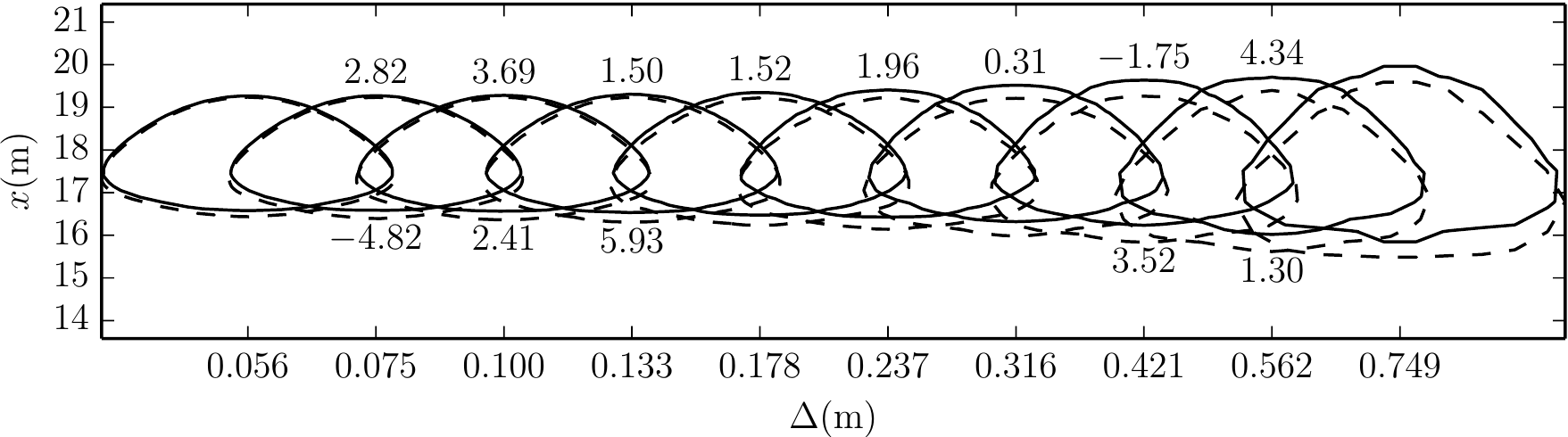}
\caption{Contour lines of the final deposition ($h=0.01\,\r{m}$) for different meshes. The cell size is shown on the horizontal axis. The vertical axis indicates global x-coordinate (top view), and the piles are presented undistorted. Solid lines show results for the Gamma scheme, and dashed lines for the upwind scheme. Numbers above piles indicate the experimental convergence order for the Gamma Scheme, and numbers below for the upwind scheme. The convergence order following Roache \citep{roache1997quantification} is calculated for the furthest point $x_{\r{max}}$ of the deposition area (i.e., the runout). Where no convergence is indicated, $x_{\r{max}}$ is locally oscillating.}
\label{fig:convergence}
\end{figure}

For a grid-convergence analysis following Roache \cite{roache1997quantification}, we consider the furthest point of the avalanche $x_{\r{max}}$, the so-called runout (determined by the $0.01\,\r{m}$ level curve). The runout is the most relevant result of such simulations in practice. Although the deposition zone is slightly inclined, the runout does not react very sensitively to the mesh refinement. We observed an average convergence order of $1.15$ for the upwind scheme and $1.96$ for the Gamma scheme. Both values are unexpectedly high, especially considering the high (diffusive) value of $\beta$. The corresponding grid convergence index (GCI) is $0.425\%$ ($x_{\r{max}} = 19.238 \pm 0.081\,\r{m}$) for the upwind scheme and $0.011\%$ ($x_{\r{max}} = 19.268 \pm 0.002\,\r{m}$) for the Gamma scheme. Overall, the upwind scheme exhibits smaller runouts and higher uncertainties than the Gamma scheme (see Fig.~\ref{fig:convergence}).

The same procedure has been conducted with regard to the time steps, using the mesh with a cell size of $0.1\,\mathrm{m}$. To obtain different time steps, the maximum Courant number has been varied between $0.25$ and $4.0$. Larger Courant numbers were unstable, and the respective simulations failed. We observed an average convergence order of $1.77$ and a corresponding GCI of $0.021\%$ ($x_{\r{max}} = 19.262 \pm 0.004\,\r{m}$). The results are summarized in Fig.~\ref{fig:convergence}. During this procedure, we observed that the computational time was almost equal for all simulations with $C_{\r{max}} \geq 1$ because larger time steps required more iterations to achieve convergence. The very small tolerance for the initial residual of $10^{-6}$, and accordingly a high number of iterations for each step (see Table~\ref{tab:params}), are necessary to attain a good convergence. We assume that the error, accumulated due to imprecise solving of the nonlinear equations, leads to higher uncertainties than the discretization error. 

\begin{figure}
\centering
\includegraphics[scale=0.67]{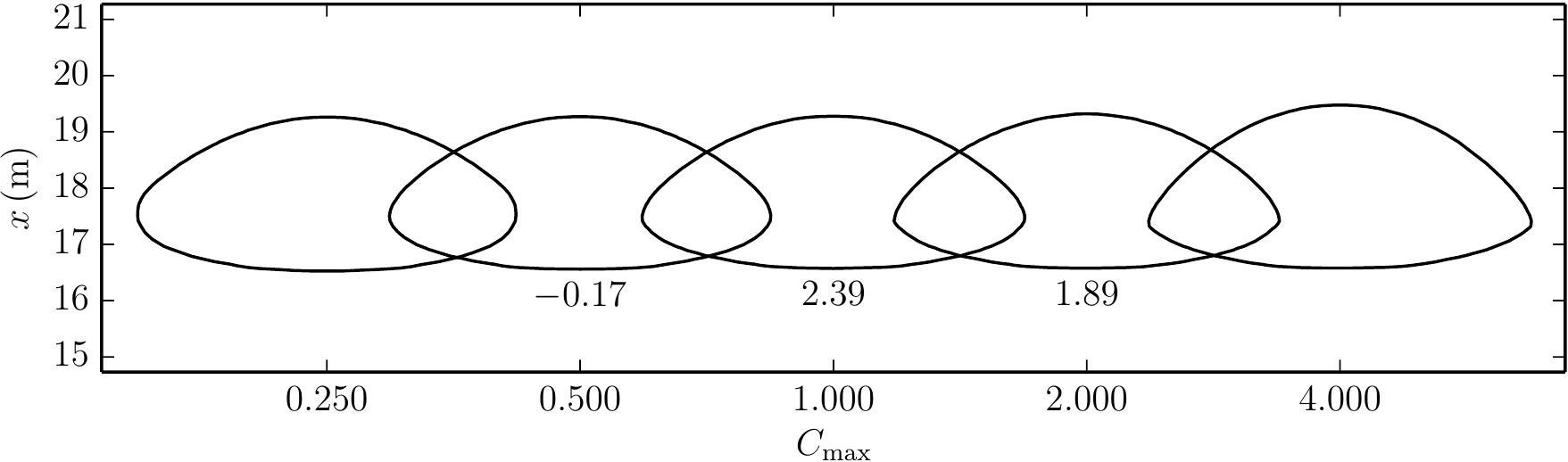}
\caption{Contour lines of the final deposition ($h=0.01\,\r{m}$) for different Courant numbers and therefore different time step durations. The maximum Courant number is shown on the horizontal axis. The vertical axis indicates the global x-coordinate (top view), and the piles are presented undistorted. Numbers below piles indicate the experimental convergence order for the time integration scheme. The convergence order \citep{roache1997quantification} is calculated for the furthest point $x_{\r{max}}$ of the deposition area (i.e., the runout). }
\label{fig:timeconvergence}
\end{figure}

\subsection{Granular avalanche on a curved and twisted slope}

\begin{figure}
\centering
\includegraphics[width=\columnwidth]{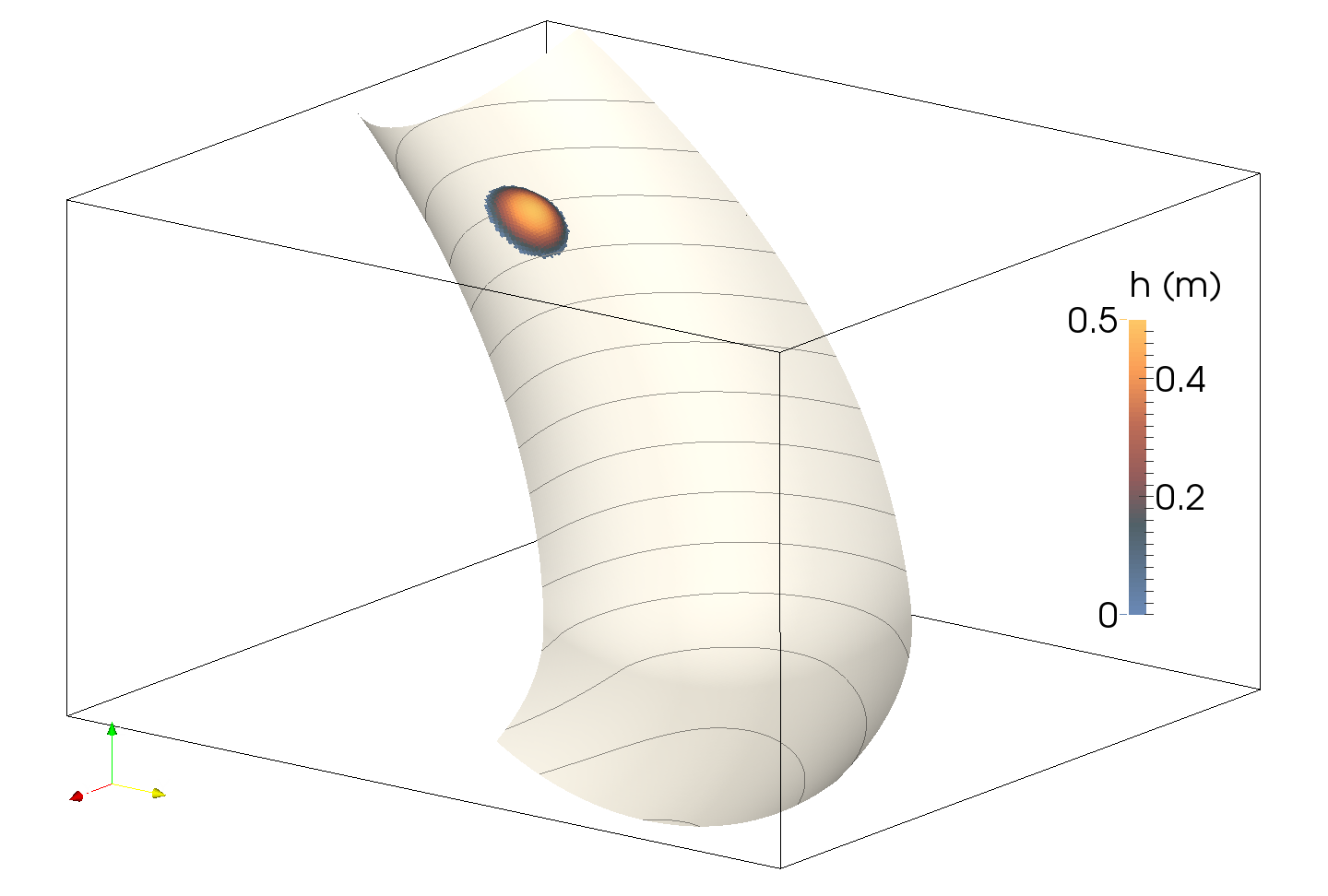}
\caption{Slope geometry and initial condition of the curved and twisted slope. The gray contour lines show the elevation in $1\,\r{m}$ steps.}
\label{fig:twisted_initial}
\end{figure}

\begin{figure}
\centering
\includegraphics[width=\columnwidth]{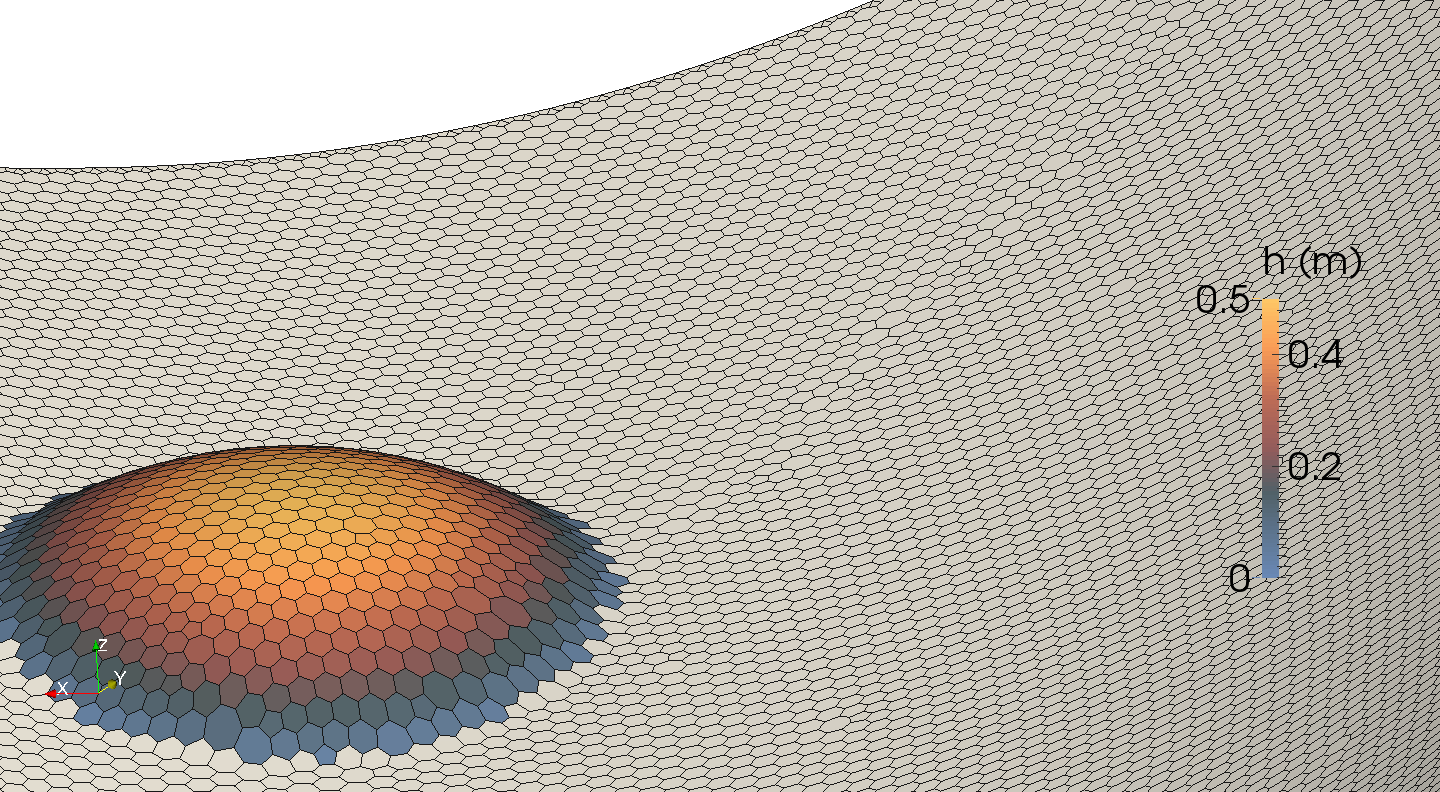}
\caption{A part of the mesh containing the initial condition of the curved and twisted slope. The colored area indicates the position of the free surface in space $\b{x}_{\r{fs}} = \b{x}_{\r{b}}-\b{n}_{\r{b}}\,h \ \forall \b{x}_{\r{b}} \in \Gamma$, and the white area indicates the basal surface $\Gamma$.}
\label{fig:twisted_mesh}
\end{figure}

Finally, we aim to present a complexly curved slope with a twisted flow path to demonstrate the capability of handling complex terrain and unstructured meshes. The surface in this example is based on the simply curved surface from section~\ref{sec:res_simply_avalanche}, but with a parabolic cross section, based on the function $\Delta \loc{z} = \loc{y}^2/7$. Moreover, the surface has been bent, such that the projection of the central flow path in the x-y-plane forms a curve with radius $20\,\r{m}$. The resulting surface is shown in Fig.~\ref{fig:twisted_initial}, and a part of the mesh is presented in Fig.~\ref{fig:twisted_mesh}. All parameters are taken from the previous example. From Fig.~\ref{fig:twisted_mesh}, it can be seen that the flow thickness is small in comparison to the curvature radius in this example, as well. The mesh, a surface with an area of $325.8\,\r{m}^{2}$, consists of $38\,146$ polygonal cells, and the average cell size follows as $0.092\,\r{m}$. The mesh has been created based on a STL file, containing a triangulation of the surface. This method can be applied to natural terrains, as well. 
Note that all terrain information is encoded in the unstructured mesh. At this stage, the twisted slope can not be considered as a special case, but rather as arbitrary topography.
In this study, we applied the mesh generator Netgen \cite{schoberl1997netgen} to create a tetrahedral mesh out of the STL surface. A polyhedral mesh (which is usually preferred over a tetrahedral mesh) has been obtained by generating the dual mesh to the tetrahedral mesh \cite[p. 147]{moukalled2016finite}. However, we obtained virtually the same results with the polyhedral and tetrahedral mesh. Note that a tetrahedral mesh guarantees flat faces, while polyhedral meshes may contain some warped faces, in which not all points are located on a single plane. A valuable indicator for this issue is the ratio between the sum of areas of individual triangles and the same area projected on a flat surface. This ratio is $>0.9999$ for every cell of the surface mesh used here, where $1$ indicates perfectly flat meshes.


\begin{figure}
\centering
\includegraphics[scale=0.75]{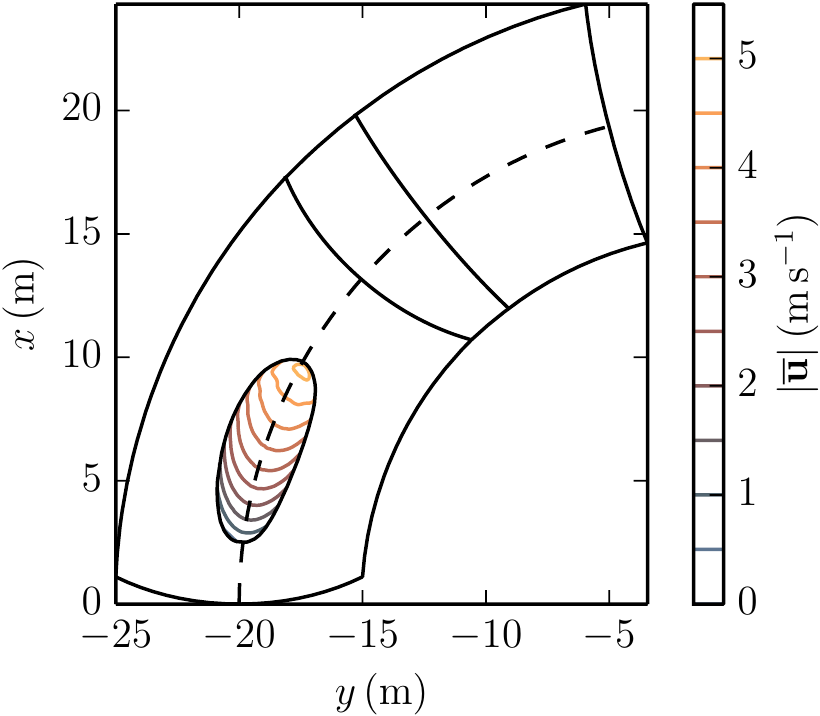}\includegraphics[scale=0.75]{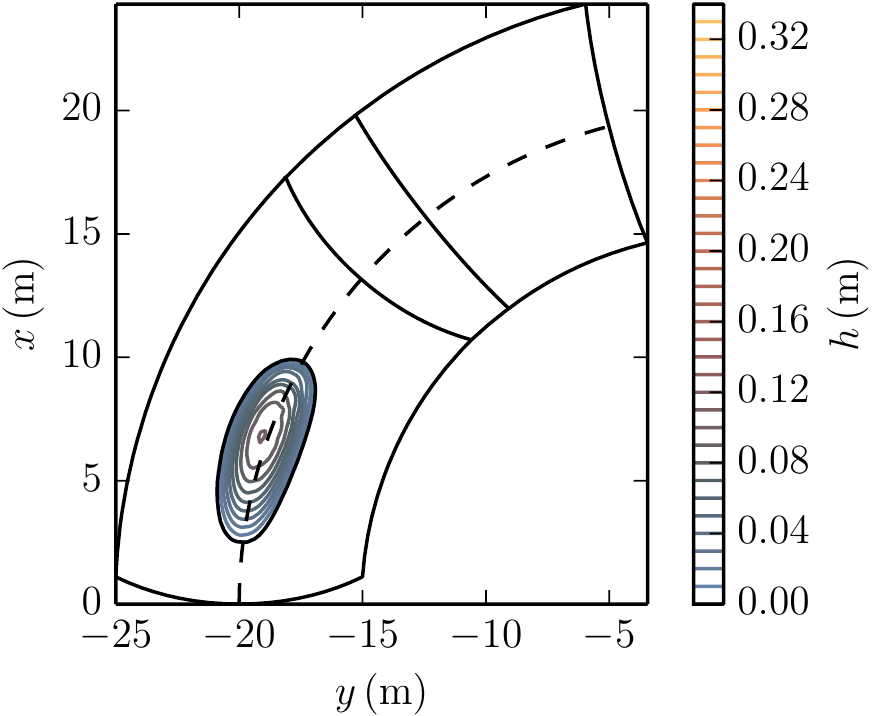}\\[3ex]
\includegraphics[scale=0.75]{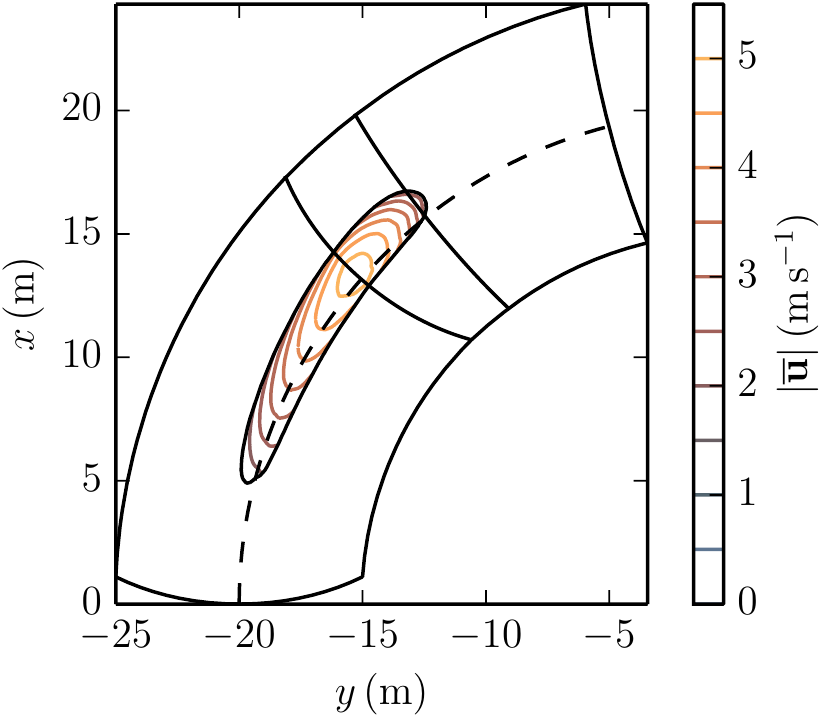}\includegraphics[scale=0.75]{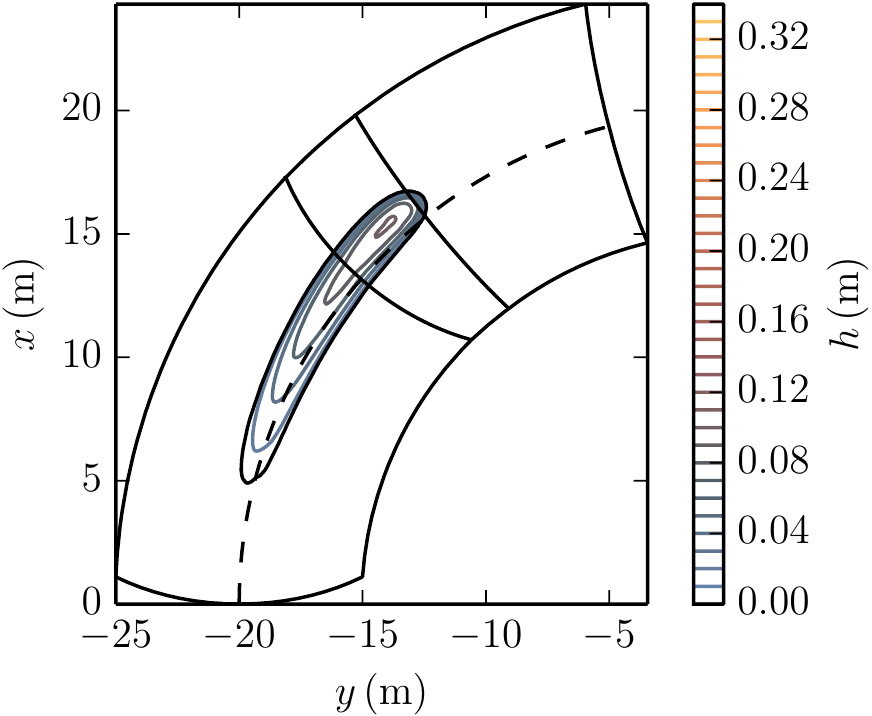}\\[3ex]
\includegraphics[scale=0.75]{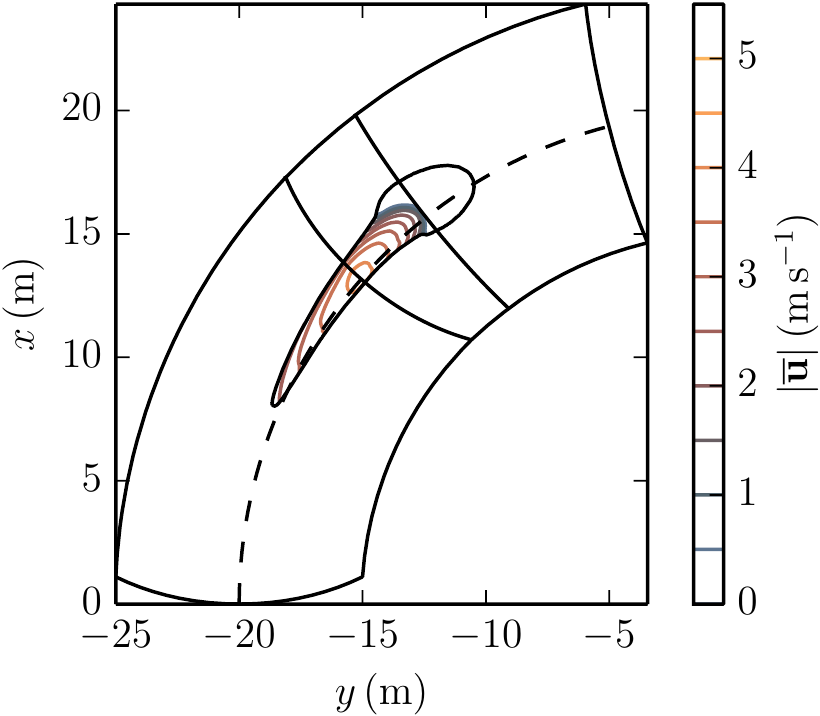}\includegraphics[scale=0.75]{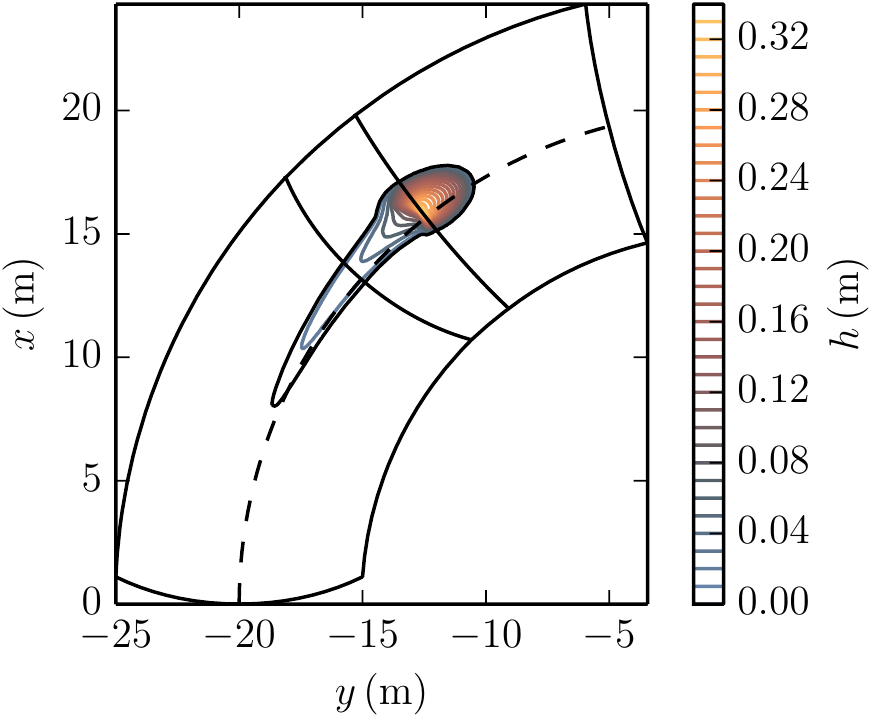}\\[3ex]
\caption{Top view of the curved and twisted slope for times $t=2\,\r{s}$ (top), $t=4\,\r{s}$ (middle), and $t=6\,\r{s}$ (bottom). Contour lines show the depth-averaged velocity in $0.5\,\r{m\,s^{-1}}$ steps (left) and the flow thickness in $0.01\,\r{m}$ steps (right). Areas in which the flow thickness is lower than $0.01\,\r{m}$ are not shown. The contour of the slope is indicated in black, and the transition zone is marked by two black lines. The middle line of the slope (e.g., a circle with radius $20\,\r{m}$ in the x-y-plane) is presented as a dashed line.}
\label{fig:curved_slope_run}
\end{figure}

The results are shown in Fig.~\ref{fig:curved_slope_run}. The avalanche approximately follows the line of steepest slope. Inertia pushes the avalanche in the orographic left direction, while basal pressure forces the avalanche back towards the middle line. The eccentricity follows from the balance of inertia and basal pressure. The pile comes to rest on the orographic left side. In comparison to the example on the simply curved slope, the lateral spreading of the avalanche is distinctly reduced. The lateral confinement leads to a higher velocity. The influence on the runout is surprisingly small, and the highest point of the final deposition is in both cases near the end of the transition zone. The obtained results qualitatively match the results of Pudasaini et al. \cite{pudasaini2005rapid}.

\section{Conclusion and outlook}
\label{sec:outlook}

This work shows the derivation, implementation, and application of a depth-integrated shallow flow model for granular materials on three-dimensional topographies. The interaction with the three-dimensional surface is considered with a novel approach. The shown model yields a simple solution for the basal pressure field, which is critical for the accurate description of granular rheology.

The proposed model has been implemented in OpenFOAM (version foam-extend-4.0). OpenFOAM provides an ideal platform for such tasks. The top level solver code, mainly mimicking the Shallow Water Equations, appears comprehensible and constitutes the ideal basis to employ the large variety of extensions originally developed for the Savage-Hutter model. The governing equations are solved directly on the boundary surface of a finite volume mesh. This simplifies the coupling of surface transport equations with three-dimensional ambient flow equations, which is important for so-called mixed and powder snow avalanches \cite{sampl2004avalanche}.

We succeeded in closing the gap to former solutions from the granular flow and thin liquid film communities. The shown implementation allows sharing a high amount of code, which is well-suited for the general surface transport problem, while keeping it adaptable to the specific parts. Considering the growing demands and the complexity of fluid dynamics codes, it seems unpromising to develop independent solutions for small communities.

To quantitatively verify the numerical routine, dam break cases and a steady flow over a constantly curved surface have been utilized. 

Simulations of granular avalanches on curved slopes demonstrate the handling of arbitrary, but mildly curved, topographies and the application of the $\mu(I)$-rheology. We applied the simplest method to include rheology, yielding solely a basal friction term. This so-called local rheology is sufficient for the case of a granular avalanche. For more complex cases, a non-local rheology (e.g., \cite[][]{baker2016two}) has to be applied.
 
The application of the implicit scheme shows some advantages, i.e., solid-like behavior in the runout zone, theoretical unconditional stability, and simple implementation, since an equation evaluation is only required at a single time level. However, many iterations are required to reach the desired residual, making the implicit method expensive in terms of computational effort. Although implicit schemes are preferred by the thin film community, their application to granular avalanches seems inefficient. This can probably be attributed to the jump of flow conditions at the moving front between dry areas and the avalanche, in which the implicit solver requires many iterations to find an equilibrium. However, the application of explicit schemes, such as the Adams-Bashforth \cite[p. 139]{ferziger2002computational} or Runge-Kutta method \citep[p. 141]{ferziger2002computational}, may lead to better performance in terms of processor time, since no iterations within a single time step are required. Moreover, no linear system of equations has to be solved. However, this will require a more complex treatment of rheology, mainly in the runout zone. A semi-implicit method, handling only the basal friction term implicitly, seems promising and will constitute the topic of our future work.

\section{Acknowledgments}

We gratefully acknowledge financial support from the OEAW project ''Beyond Dense Flow Avalanches''. The computational results presented have been achieved (in part) using the HPC infrastructure LEO of the University of Innsbruck.\\
M. Rauter thanks Hrvoje Jasak, Henrik Rusche, and Vuko Vuk\v{c}evi{\'c} for many valuable comments and training regarding OpenFOAM, as well as all supervisors and fellows at the University of Innsbruck and the Federal Research Centre for Forest for continuous encouragement and valuable discussions.
Moreover, we are grateful to Alexander Ostermann and Dieter Issler for assistance regarding the time integration scheme and mathematical notations.
We thank the reviewers and editors for valuable comments which improved the clarity and quality of this paper substantially.

\bibliography{fam.bib}

\end{document}